# Comparison of g-estimation approaches for handling symptomatic medication at multiple timepoints in Alzheimer's Disease with a hypothetical strategy

Lasch, Florian[1,2]* Guizzaro, Lorenzo[1,3]*, Loh, Wen Wei[4]

1. European Medicines Agency, Amsterdam, The Netherlands.

2. Hannover Medical School, Carl-Neuberg-Straße 1, 30625 Hannover, Germany

3. Medical Statistics Unit, Department of Physical and Mental Health and Preventive Medicine, University of Campania "Luigi Vanvitelli", Naples, Italy

4. Department of Methodology & Statistics, Faculty of Health, Medicine and Life Sciences (FHML), Maastricht University, Maastricht, The Netherlands.

* These authors have contributed equally to this work; order was assigned by tossing a coin.

## Disclaimer

The views expressed in this article are the personal views of the author(s) and may not be understood or quoted as being made on behalf of or reflecting the position of the European Medicines Agency or one of its committees or working parties.

## Funding

No funding was received for this study.

## Disclosure statement

*The authors report there are no competing interests to declare.*

**Abstract**

For handling intercurrent events in clinical trials, one of the strategies outlined in the ICH E9(R1) addendum targets the hypothetical scenario of non-occurrence of the intercurrent event. While this strategy is often implemented by setting data after the intercurrent event to 'missing' even if they have been collected, g-estimation allows for a more efficient estimation by using the information contained in post-IE data. As the g-estimation methods have largely developed outside of randomised clinical trials, optimisations for the application in clinical trials are possible. In this work, we describe and investigate the performance of modifications to the established g-estimation methods, leveraging the assumption that some intercurrent events are expected to have the same impact on the outcome regardless of the timing of their occurrence. In a simulation study in Alzheimer's disease, the modifications show a substantial efficiency advantage for the estimation of an estimand that applies the hypothetical strategy to the use of symptomatic treatment while retaining unbiasedness and adequate type I error control.

## 1. Introduction

The introduction of the estimand framework [1] has allowed framing precise research questions to guide the design and analysis of randomized clinical trials. The main novelty of this framework is the consideration of intercurrent events (IEs) – defined as events occurring after randomization that affect the interpretation of the variable of interest or preclude its observation. The ICH E9 (R1) addendum on estimands and sensitivity analysis [1] describes different strategies to handle IEs, each strategy reflecting a different underlying research question of interest and resulting in a different estimand. One of the five strategies described for handling IEs is the hypothetical strategy, reflecting the interest in the effect of a medicine in case an intercurrent event had not occurred.

A specific instance where an estimand that uses a hypothetical strategy may be of interest is for the use of medications – such as acetylcholinesterase inhibitors or memantine - that produce a cognitive improvement in absence of a substantial effect on the underlying pathology (hereinafter referred to as "symptomatic medications") in a trial of a disease-modifying treatment in early Alzheimer's Disease [2]. This motivating example will be used throughout the article for comparing the performance of different estimators for an estimand using the hypothetical strategy, described in more detail below. While the clinical decision to initiate symptomatic treatment does not depend on a single outcome measured in a patient, the probability to initiate symptomatic medication is proportional to the observed decline. Therefore, in trials for a truly disease-modifying treatment, symptomatic medication will be initiated more often - in expectation - in the placebo arm as compared to the active treatment arm, potentially reducing the observed effect of the disease modifying treatment if not accounted for. Correspondingly, for this type of trials, the clinical interest can be in estimating the difference in outcomes between patients assigned to a candidate disease-modifying medicine and to placebo, without the effect of symptomatic medications. In trials of long durations – such as those generally conducted for candidate disease-modifying medicines [3, 4] – it is reasonable to assume that the initiation of symptomatic treatments – our IE of interest – can occur at different timepoints. The main question of this article – motivated by the EMA guideline on Alzheimer's Disease [2] - will be on what the best estimation approach is for the above-described situation.

The most widespread approach for estimating estimands that uses a hypothetical strategy has been to ignore data that arise after the IE, by treating them as missing data, and then use missing-data methods such as Multiple Imputation [5], Inverse Probability Weighting [6], or Mixed Effects Modelling [7]. These methods do not use the information that is encoded in the post-IE measurements. However, using such information might be advantageous. We have shown in previous work that there are situations where g-estimation methods that de-mediate the effect of the IE from the observed values, in particular the g-estimation approach [8, 9], can outperform the approaches based on missing-data methods [10, 11]. In this context, it is important to highlight that the proposed de-mediation approach does not contrast with the recommendation not to include covariates measured after baseline in the main analysis [12]. Instead, we suggest that such recommendation should be read as regarding the main analysis model, and not referring to including the occurrence of intercurrent events in models used for imputation or de-mediation.

In our previous work, for simplicity, we have used a data-generating mechanisms where the IE could only occur at one intermediate timepoint. Scenarios where the same IE can occur at several intermediate timepoints during a clinical trial are realistic and pose a non-trivial question of how to best implement the de-mediation approaches. In this paper, we will focus on such common and realistic scenarios.

In the context of the types of trials and events considered in this paper, it can be reasonably further assumed that the effect of the occurrence of the intercurrent event, in our case initiating symptomatic mediation, on the outcome is the same regardless of the timepoint of occurrence. However, existing de-mediation methods focus on the different effects of the mediators at each timepoint one-at-a-time and do not readily permit fixing the effects to be the same at multiple time points. The objective of this paper is to explore and compare different proposed extensions of the established longitudinal de-mediation approach [8] for the estimation of an estimand that uses a hypothetical strategy for a single intercurrent event that can occur at multiple timepoints.

## 2. Methods

In the following sections, we first introduce the clinical trial setting, relevant variables and causal structure as a basis to then define the estimand of interest. Thereafter, we introduce in some detail the different g-estimation implementations for targeting this estimand. Subsequently, we describe the design of a simulation study based on the clinical trial setting and causal structure outlined before that investigates the performance of the g-estimation approaches. Lastly, we provide a description of the performance criteria used for comparing the approaches.

### *2.1. Trial setting and causal structure*

As a motivating starting point, we consider a randomized clinical trial (RCT) in Early Alzheimer's Disease where a disease-modifying investigational treatment is compared with placebo at 2 years, a common duration for Alzheimer's trials [13, 14], using as primary endpoint the ADAS-Cog 13 scale [15]. ADAS-Cog 13 is a scale that measures memory and other cognitive functions and has scores ranging from 0 to 85 (higher scores indicate greater impairment). For this case study, we assume that the disease-modifying treatment – our IE of interest – is administered at regular intervals for the whole duration of the trial and that the primary endpoint ($Y_t$) is measured at baseline and four consecutive timepoints $t \in \{0.5, 1, 1.5, 2\}$.

The relevant variables to describe the clinical setting are:

$t \in \{0.5, 1, 1.5, 2\}$: timepoint of measurement in years;

$Y_t^*$: The counterfactual value of ADAS-Cog13 would have at time $t$ if patients had not started symptomatic treatment before $t$ (potentially unobserved);

$Y_t$: Observed value of ADAS-Cog13 at time $t$, that may differ from $Y_t^*$ for patients who initiated symptomatic treatment;

*prognosis*: an unobserved patient characteristic that influences both baseline severity and the *decline*;

*decline*: an unobserved patient characteristic that influences the expected deterioration of symptoms during the trial;

$Z$: Observed randomized treatment assignment to placebo or to the investigational disease-modifying treatment;

$Sym_t$: initiating symptomatic treatment at timepoint $t$. We assume that $Sym_t$ is always observed with symptomatic treatment possibly started at $t \in \{0.5, 1, 1.5\}$ and that it is continued from the initiation until the end of the study. As this variable encodes initiation and not exposure, it can be 1 at only one timepoint for each simulated patient. As we do not model discontinuation of symptomatic treatment, initiating it at timepoint $t$ will affect all subsequent $Y_{t'}$ at later times t'>t.

The causal relationships that we assume among the variables are visually illustrated in the causal directed acyclic graph (DAG) in Figure 1. Causal DAGs are comprehensively described elsewhere in the literature [16-18]. As a short summary, causal DAGs consist of two elements: nodes (corresponding to variables) and arrows (representing a causal effect of one variable on another). We adopt the convention that each variable is represented by a square (if observed) or by an ellipse (if unobserved or potentially unobserved). Variables are linked by arrows that indicate the possibility of a causal relationship based on theoretical knowledge: the variable at the beginning of the arrow may exert a direct causal effect on the variable at the end of the arrow. In contrast, only when it can be justified that no direct causal effect between two variables exists is there no arrow connecting them.

*Figure 1: A DAG encoding the causal structure of the data-generating model.*

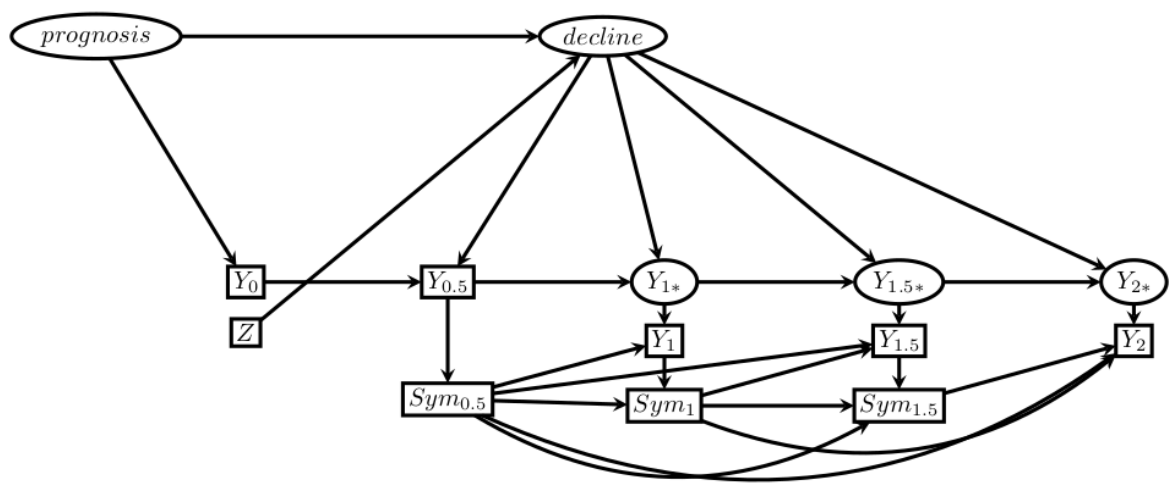

The DAG above describes the causal structure of the Monte Carlo simulations later in this paper. We will simulate trials under the null hypothesis of no direct effect (to be formally defined later), which is described by the same causal DAG without the arrow from Z to decline, as well as under the alternative hypothesis. A detailed description of the DGM follows in the section 'simulation overview'. Below we highlight a few aspects about the symptomatic effects, from our causal DAG and data generating mechanism (described in more detail below), that differ from the longitudinal de-mediation approach of Loh et al [8]:

1. The symptomatic effect (defined as the $Sym_t \to Y_{t+0.5}$ arrow) does not affect the underlying disease severity $Y_t^*$, but just the observed outcome $Y_t$. Here, we adopted this causal structure because of the mode of action of a symptomatic medication and to distinguish it from the disease-modifying effect of the investigational treatment.

2. As previously stated in the Introduction, we will assume that the expected effect of the symptomatic treatment on the observed outcome, $Sym_{t_1} \to Y_{t_2}$, is the same across all time points $t_1 < t_2$. In particular, we assume that the onset of the full effect of the mediator happens before the next consecutive timepoint and that there is no reduction in effect over time.

### 2.2. Clinical question of interest and formal definition of the estimand

The clinical question of interest in our motivating example is whether an investigational medicine slows down progression in subjects with Prodromal AD. In this setting, we consider one relevant intercurrent event, the initiation of a symptomatic medication, for which a hypothetical strategy is used. Taken together, this results in the following detailed definition of the estimand of interest:

**treatment condition of interest**: Investigational disease-modifying treatment without use of symptomatic medications;

**alternative treatment condition**: Placebo without use of symptomatic medications;

**population**: Patients with Prodromal AD;

**variable**: change in the ADAS-Cog13 after 24 months as compared to baseline;

**population-level summary**: Difference in mean change from baseline in the ADAS-Cog 13 scale after 24 months.

**Intercurrent event and strategy for handling**: Initiation of symptomatic treatment - hypothetical strategy (as if the IE - i.e. the initiation of symptomatic treatment - had not taken place).

This estimand corresponds to the Controlled Direct Effect (CDE) [19] of the investigational disease-modifying treatment with no use of symptomatic treatments.

### 2.3. Estimators

Before describing the proposed estimators, we first determine the covariates sufficient to satisfy the sequential ignorability assumptions for nonparametrically identifying the CDE estimand. The assumption states that each mediator-outcome relationship (e.g., the symptomatic effect $Sym_t \to Y_{t+0.5}$) can be d-separated; see assumption (A2) of Loh et al. (2020). Hence, we used dagitty [20] to recreate the causal diagram in Figure 1 (https://dagitty.net/mowd92rkA). By setting $Sym_t$ as the "exposure" and $Y_s, s > t$ as the "outcome" in dagitty, we can determine a minimal sufficient adjustment set for each mediator-outcome relationship. For example, the adjustment set for the $Sym_{1.5} \to Y_2$ effect is $(Sym_{0.5}, Sym_1, Y_{1.5})$. Below we list the adjustment sets for all the effects (Table 1).

*Table 1: minimal adjustment sets for different effects in the causal DAG*

| Effect to be estimated | Minimal adjustment set |
|---|---|
| $Sym_{1.5} \to Y_2$ | $Sym_{0.5}, Sym_1, Y_{1.5}$ |
| $Sym_1 \to Y_2$ | $Sym_{0.5}, Y_1$ |
| $Sym_1 \to Y_{1.5}$ | $Sym_{0.5}, Y_1$ |
| $Sym_{0.5} \to Y_2$ | $Y_{0.5}$ |
| $Sym_{0.5} \to Y_{1.5}$ | $Y_{0.5}$ |
| $Sym_{0.5} \to Y_1$ | $Y_{0.5}$ |

In the following section, we describe the different implementations of g-estimation methods that we will compare in this simulation study.

#### 2.3.1. Established g-estimation for longitudinal de-mediation

Loh and colleagues have proposed a g-estimation method for longitudinal de-mediation for estimating the controlled direct effect, by sequentially de-mediating the effect of the mediator at each time point on an end-of-study outcome starting with the latest timepoint [8].

The above described estimand of interest and causal structure, with the same mediator affecting the outcome at multiple timepoints, is a special case of the more general context studied by Loh and colleagues, and therefore their method is expected to be unbiased. Applied to our case study, where we target the CDE of Z on Y2 fixing all $Sym_t$ at 0, the longitudinal de-mediation approach by Loh et al follows the following steps:

i. define the variable $R_2$ - that will be used to de-mediate the impact of symptomatic treatment from the observed outcome in reverse time-order - as $R_2 = Y_2$.

ii. For each timepoint where the observed values can be affected by the symptomatic treatment intake $t \in \{2, 1.5, 1\}$, starting with the latest timepoint and moving to the earlier timepoints, the following de-mediation steps are conducted:

### de-mediating $Sym_{1.5} \to Y_2$

a. Predict the probability for having started symptomatic treatment at $t = 1.5$ for all patients by fitting a logistic model using the "probit" link function $P(Sym_{1.5} = 1) \sim Sym_1 + Sym_{0.5} + Y_{1.5}$ and using this model to estimate the probability to start symptomatic treatment $p_{sym,i}$ for all patients $i$.

b. Fit a linear model $R_2 \sim Z + Y_{1.5} + Sym_{1.5} + Sym_1 + Sym_{0.5} + p_{sym}$. The coefficient of $Sym_{1.5}$ from this model, $coef(Sym_{1.5})$, estimates the effect $Sym_{1.5} \to Y_2$.

c. De-mediate the effect of $Sym_{1.5}$ via $R_{1.5,i} = R_{2,i} - coef(Sym_{1.5}) * Sym_{1.5,i}$ using the coefficient of $Sym$ from step (b);

### de-mediating $Sym_1 \to Y_2$

a. Predict the probability for having started symptomatic treatment at $t = 1$ for all patients by fitting a logistic model using the "probit" link function $P(Sym_1 = 1) \sim Sym_{0.5} + Y_1$ and using this model to estimate the probability to start symptomatic treatment $p_{sym,i}$ for all patients $i$.

b. Fit a linear model $R_{1.5} \sim Z + Y_1 + Sym_1 + Sym_{0.5} + p_{sym}$. The coefficient of $Sym_1$ from this model, $coef(Sym_1)$, estimates the effect $Sym_1 \to Y_{1.5}$, and given our DGM also provides an estimate of the direct effect of initiating symptomatic treatment at t=1 on the final outcome (i.e. the effect $Sym_1 \to Y_2$ that does not intersect $Sym_{1.5}$).

c. De-mediate the effect of $Sym_1$ via $R_{1,i} = R_{1.5,i} - coef(Sym_1) * Sym_{1,i}$ using the coefficient of $Sym$ from step (b);

### de-mediating $Sym_{0.5} \to Y_2$

a. Predict the probability for having started symptomatic treatment at $t = 0.5$ for all patients by fitting a logistic model using the "probit" link function $P(Sym_{0.5} = 1) \sim Y_{0.5}$ and using this model to estimate the probability to start symptomatic treatment $p_{sym,i}$ for all patients $i$.

b. Fit a linear model $R_1 \sim Z + Y_{0.5} + Sym_{0.5} + p_{sym}$. The coefficient of $Sym_{0.5}$ from this model, $coef(Sym_{0.5})$, estimates the effect $Sym_{0.5} \to Y_1$, and given our DGM also provides an estimate of the direct effect of initiating symptomatic treatment at t=0.5 on the final outcome.

c. De-mediate the effect of $Sym_{0.5}$ via $R_{0.5,i} = R_{1.5,i} - coef(Sym_{0.5}) * Sym_{0.5,i}$ using the coefficient of $Sym$ from step (b);

iii. Estimate the direct effect of the treatment via the linear model $R_{0.5} \sim Z + Y_0$. While the inclusion of $Y_0$ is not necessary for unbiased estimation, it is expected to increase the precision of the treatment effect estimate.

The two key assumptions on this method are that there are (i) no unobserved confounder between treatment and outcome and (ii) no unobserved confounders between the mediator and the outcome. The former is satisfied because our treatment Z is randomized, and the latter is satisfied under the causal DAG in Figure 1 above using dagitty.

In step (ii) above, the established g-estimation approach for longitudinal de-mediation is carried out iteratively for the mediator at three separate time points. However, in our case study, we assume that the mediator only affects the observed outcome (instead of the underlying disease severity) and that the same mediator (the initiation of symptomatic medication) can occur at multiple timepoints with the same expected effect. Thereby, in the following, we are describing two modifications to the established approach that aim to utilize the fact that the same mediator occurs at multiple time points. With reference to the causal DAG, based on the fact that the same mediator can occur at any of the three time points, we will hereafter consider the realistic assumptions that (a) the mediator effects from $Sym_{t-0.5}$ on $Y_t$ are the same in expectation and (b) the mediator effects from $Sym_{t-0.5}$ on $Y_2$, assuming no effect of $Sym_s$ at intermediate time points $t - 0.5 < s < 2$, are the same in expectation for all t. Based on these assumptions, the established g-estimation approach for longitudinal de-mediation can be modified to benefit from assuming equal mediator effects.

In the following, we propose modifications based on assumptions (a) and (b) above respectively using the same underlying idea: instead of de-mediating the effect of $Sym_t$ at each time point separately, as done by the established g-estimation approach for longitudinal de-mediation, the effects of $Sym_t$ (either on $Y_{t-0.5}$ or $Y_2$) can be averaged over all t. Our proposals are based on the average being a more precise estimate of the symptomatic effect for de-mediation.

### 2.3.2. Modification 1 - Averaging the de-mediation effects $Sym_{t-0.5}$ on $Y_t$

Given that the mediator has the same average effect regardless of the timepoint of initiation, averaging the individual estimates of the symptomatic effects from each $Sym_{t-0.5}$ to each $Y_t$ is expected to estimate the de-mediation effect more precisely. For doing so, modification 1 of the established g-estimation approach for longitudinal de-mediation follows the subsequent steps:

**estimating $Sym_{0.5} \rightarrow Y_1$**

a. Predict the probability for having started symptomatic treatment at $t = 0.5$ by fitting a logistic model with the "probit" link function $P(Sym_{0.5} = 1) \sim Y_{0.5}$ and using this model to estimate the probability to start symptomatic treatment $p_{sym,i}$ for all patients $i$.

b. Fit a linear model $Y_1 \sim Z + Y_{0.5} + Sym_{0.5} + p_{sym}$. The coefficient of $Sym_{0.5}$ from this model, $coef(Sym_{0.5})$, estimates the effect $Sym_{0.5} \rightarrow Y_1$.

**estimating $Sym_1 \rightarrow Y_{1.5}$**

a. Predict the probability for having started symptomatic treatment at $t = 1$ by fitting a logistic model with the "probit" link function $P(Sym_1 = 1) \sim Y_1 + Sym_{0.5}$ and, using this model, predicting the probability to start symptomatic treatment $p_{sym,i}$ for all patients.

b. Fit a linear model $Y_{1.5} \sim Z + Y_1 + Sym_1 + Sym_{0.5} + p_{sym}$, restricting the analysis set to patients with $Sym_{0.5} = 0$. The coefficient of $Sym_1$ from this model, $coef(Sym_1)$, estimates the effect $Sym_1 \rightarrow Y_{1.5}$.

**estimating $Sym_{1.5} \rightarrow Y_2$**

a. Predict the probability for having started symptomatic treatment at $t = 1.5$ by fitting a logistic model with the "probit" link function $P(Sym_{1.5} = 1) \sim Y_{1.5} + Sym_1$ and, using this model, predicting the probability to start symptomatic treatment $p_{sym,i}$ for all patients.

b. Fit a linear model $Y_2 \sim Z + Y_{1.5} + Sym_{1.5} + Sym_1 + p_{sym}$. The coefficient of $Sym_{1.5}$ from this model, $coef(Sym_{1.5})$ estimates the effect $Sym_{1.5} \rightarrow Y_2$.

iii. Calculate the overall effect of initiating symptomatic medication $Sym_{t-0.5}$ on the outcome $Y_t$, $\beta_{sym}$, as the average of $coef(Sym_{t-0.5}), t \in \{2, 1.5, 1\}$, weighted by the standard error from the models in steps (b) above.

$$\beta_{sym} = \left( \frac{coef(Sym_{0.5})}{SE(coef(Sym_{0.5}))} + \frac{coef(Sym_1)}{SE(coef(Sym_1))} + \frac{coef(Sym_{1.5})}{SE(coef(Sym_{1.5}))} \right) * (SE(coef(Sym_{0.5})) + SE(coef(Sym_1)) + SE(coef(Sym_{1.5}))).$$

iv. De-mediate the joint effects of all previous symptomatic medications using the averaged effect $\beta_{sym}$ from the outcome $Y_2$ via $R = Y_2 - (Sym_{0.5} + Sym_1 + Sym_{1.5}) * \beta_{sym}$

v. Estimate the direct effect of the treatment by applying a linear regression model $R \sim Z + Y_0$.

Modification 1 differs from the original approach by deriving an overall estimate of the symptomatic effect ($\beta_{sym}$) from the individual estimates ($Sym_{1.5} \to Y_2$, $Sym_1 \to Y_{1.5}$, and $Sym_{0.5} \to Y_1$). The effects are then de-mediated concurrently in step (iv), rather than iteratively as in the original approach. We can use this method based on the assumptions that each $Sym_t \to Y_{t+0.5}$ equals $Sym_t \to Y_2$ and that the timepoint of initiation does not affect the effect of the symptomatic treatment on $Y_2$. (Assumption 2 above)

### 2.3.3. Modification 2 - Averaging the de-mediation effects $Sym_{t-0.5}$ on $Y_2$

In contrast to modification 1, which averages the mediator effects of $Sym_{t-0.5}$ on $Y_t$, the motivating idea of the second modification is to average the mediator effects directly on the outcome of interest, $Sym_{t-0.5}$ on $Y_2$, and then again utilise the assumption of equal de-mediation effects and gain a higher precision of the estimate for the de-mediation effect. In contrast to modification 1, for modification 2 backwards de-mediation is necessary since the effects of the mediators on $Y_2$ (instead of $Y_t$) are estimated.

Therefore, we propose an approach that follows the de-mediation in steps (i) and (ii) of the above-described established g-estimation approach for longitudinal de-mediation for estimating the effect of the mediator at each time point. In analogy to step (iii) in modification 1, however, these estimates are averaged before using them for de-mediating $Y_2$ and estimating the treatment effect.

For doing so, modification 2 follows steps (i) and (ii) of the established g-estimation approach for longitudinal de-mediation and thereafter averages the mediator effects $\boldsymbol{Sym_{t-0.5} \to Y_2}$ and de-mediates this average before estimating the effect $\boldsymbol{Z}$ on the de-mediated outcome.

In step (iii), we then calculate the overall effect of initiating symptomatic medication, $\beta_{sym}$, as the average of the coefficients $coef(Sym_{t-0.5})$ for $t \in \{1, 1.5, 2\}$ weighted by the standard errors (using the same formula in step (iii) of Modification 1 above) from the models in steps (ii b) of the original approach. In steps (iv) and (v), equivalently to the modification 1, we de-mediate the averaged effect $\beta_{sym}$ from the outcome $Y_2$ via $R = Y_2 - (Sym_1 + Sym_1 + Sym_{1.5}) * \beta_{sym}$ and estimate the direct effect of the treatment by applying a linear regression model $R \sim Z + Y_0$.

Modification 2 differs from the original approach in one aspect. Rather than use $R_{0.5}$ following the completion of step (ii) to estimate the direct effect, here the effects of each $Sym_{t-0.5} \to Y_2, t \in \{2, 1.5, 1\}$, are first averaged in step (iii) and then used to de-mediate all three effects concurrently in step (iv). This builds on Assumption 2 above.

### 2.3.4. Modification 3 - Iterative averaging the de-mediation effects $Sym_{t-0.5}$ on $Y_2$

The methods described above gain precision in the final estimate of the effect of the mediator by averaging different estimates. However, the benefit of this precision is only used in the final step, and not in the intermediate ones. In this extension of modification 2 we aim to incorporate the information about the mediation effect at a given timepoint not only in the end by averaging the mediator effects, but already in the de-mediation process.

Therefore, we propose an iterative approach that follows the de-mediation loop of steps (i) to (iii) of the above-described established g-estimation approach for longitudinal de-mediation multiple times, with one crucial change. At each time point, the estimates for the mediator effects from the other timepoints are incorporated as well. Since no closed form for this approach is available, we implement an iterative approach that repeatedly follows the de-mediation steps until a convergence criterion for the estimated treatment effect is met.

Concretely, and building on the description of the established g-estimation approach above, we take the parameters estimated in step (ii) as starting estimates for the first loop that will be further refined in subsequent loops in the following way.

For estimating the symptomatic effect for a $t \in \{1, 1.5, 2\}$, we can iteratively use the estimated effects from the respective two other timepoints from the previous loop to increase the precision of the estimates. Hereby we notate the effect estimate of the effect of Sym at timepoint $t$ from iteration $j$ as $est(Sym_t)_j$ and loop over the steps below until the difference between the treatment effect estimates between two consecutive loops is lower than 0.0001 or the number of iteration reaches 25.

In implementing this method, we first complete steps (i) and (ii) from the established g-estimation approach above, providing us with estimates $coef(Sym_{0.5})_1$, $coef(Sym_1)_1$ and $coef(Sym_{1.5})_1$.

Thereafter, we iterate the following steps, with $j$ notating the $j$-th iteration:

> Define the variable $R$ that is used to de-mediate the impact of symptomatic treatment from the observed outcome as $R = Y_2$.
>
> **de-mediating $Sym_{1.5} \to Y_2$**

a. Predict the probability for having started symptomatic treatment at $t = 1.5$ for all patients by fitting a logistic model using the "probit" link function $P(Sym_{1.5} = 1) \sim Sym_1 + Sym_{0.5} + Y_{1.5}$ and using this model to estimate the probability to start symptomatic treatment $p_{sym,i}$ for all patients $i$.

b. Fit a linear model $R_2 \sim Z + Y_{1.5} + Sym_{1.5} + Sym_1 + Sym_{0.5} + p_{sym}$. The coefficient of $Sym_{1.5}$ from this model, $coef(Sym_{1.5})_j$, estimates the effect $Sym_{1.5} \rightarrow Y_2$.

c. Calculate the weighted average of $coef(Sym_{1.5})_j$ and the estimates from the previous iteration of the symptomatic effect, $coef(Sym_{0.5})_{j-1}$ and $coef(Sym_1)_{j-1}$, as

$$average(Sym_{1.5})_j = \left(\frac{coef(Sym_{0.5})_{j-1}}{SE(coef(Sym_{0.5})_{j-1})} + \frac{coef(Sym_1)_{j-1}}{SE(coef(Sym_1)_{j-1})} + \frac{coef(Sym_{1.5})_j}{SE(coef(Sym_{1.5})_j)}\right) * (SE(coef(Sym_{0.5})_{j-1}) + SE(coef(Sym_1)_{j-1}) + SE(coef(Sym_{1.5})_j)).$$

d. De-mediate the effect of $Sym_{1.5}$ via $R_{1.5,i} = R_{2,i} - average(Sym_{1.5})_j * Sym_{1.5,i}$;

**de-mediating $Sym_1 \rightarrow Y_2$**

a. Predict the probability for having started symptomatic treatment at $t = 1$ for all patients by fitting a logistic model using the "probit" link function $P(Sym_1 = 1) \sim Sym_{0.5} + Y_1$ and using this model to estimate the probability to start symptomatic treatment $p_{sym,i}$ for all patients $i$.

b. Fit a linear model $R_{1.5} \sim Z + Y_1 + Sym_1 + Sym_{0.5} + p_{sym\prime}$. The coefficient of $Sym_1$ from this model, $coef(Sym_1)$, estimates the effect $Sym_1 \rightarrow Y_2$ that does not intersect $Sym_{1.5}$.

c. Calculate the weighted average of $coef(Sym_1)_j$ and the estimates from the previous iteration of the symptomatic effect, $coef(Sym_{0.5})_{j-1}$ and $coef(Sym_{1.5})_{j-1}$, as

$$average(Sym_1)_j = \left(\frac{coef(Sym_{0.5})_{j-1}}{SE(coef(Sym_{0.5})_{j-1})} + \frac{coef(Sym_1)_j}{SE(coef(Sym_1)_j)} + \frac{coef(Sym_{1.5})_{j-1}}{SE(coef(Sym_{1.5})_{j-1})}\right) * (SE(coef(Sym_{0.5})_{j-1}) + SE(coef(Sym_1)_j) + SE(coef(Sym_{1.5})_{j-1})).$$

d. De-mediate the effect of $Sym_1$ via $R_{1,i} = R_{1.5,i} - average(Sym_1)_j * Sym_{1,i}$;

**de-mediating $Sym_{0.5} \rightarrow Y_2$**

a. Predict the probability for having started symptomatic treatment at $t = 0.5$ for all patients by fitting a logistic model using the "probit" link function $P(Sym_{0.5} = 1) \sim Y_{0.5}$ and using this model to estimate the probability to start symptomatic treatment $p_{sym,i}$ for all patients $i$.

b. Fit a linear model $R_1 \sim Z + Y_{0.5} + Sym_{0.5} + p_{sym}$, the coefficient of $Sym_{0.5}$ from this model, $coef(Sym_{0.5})$ estimates the effect $Sym_{0.5} \rightarrow Y_2$ that does not intersect $Sym_{1.5}$ or $Sym_1$.

c. Calculate the weighted average of $coef(Sym_{0.5})_j$ and the estimates from the previous iteration of the symptomatic effect, $coef(Sym_1)_{j-1}$ and $coef(Sym_{1.5})_{j-1}$, as

$$average(Sym_{0.5})_j = \left( \frac{coef(Sym_{0.5})_j}{SE(coef(Sym_{0.5})_j)} + \frac{coef(Sym_1)_{j-1}}{SE(coef(Sym_1)_{j-1})} + \frac{coef(Sym_{1.5})_{j-1}}{SE(coef(Sym_{1.5})_{j-1})} \right) * (SE(coef(Sym_{0.5})_j) +$$

$$SE(coef(Sym_1)_{j-1}) + SE(coef(Sym_{1.5})_{j-1})).$$

d. De-mediate the effect of $Sym_{0.5}$ via $R_{0.5,i} = R_{1.5,i} - average(Sym_{0.5})_j * Sym_{0.5,i}$ ;

iv. Estimate the direct effect of the treatment via the linear model $R_{0.5} \sim Z + Y_0$ as $coef(Z)_j$

v. Stop the iteration if either j reaches 25 or $|coef(Z)_j - coef(Z)_{j-1}| < 0.0001$.

The estimate from the last iteration j, $coef(Z)_j$, is the estimate for the direct effect of the disease modifying treatment.

### 2.3.5. Standard errors

For all the three methods described above, to derive estimates for the standard errors, confidence intervals and statistical significance of hypothesis testing, we implemented three approaches: as recommended by Loh et al. [19], we applied bootstrap using on the boot function from the R package boot [20, 21] with the "basic" method. In addition, we implemented the jackknife algorithm for estimating the standard errors [21]. Lastly, we used the model-derived standard error from the last estimation step in each approach.

### 2.3.6. Reference methods

As reference methods in our simulation study, we apply the commonly used {Ting, 2023 #12} approach for analyzing these types of trials of setting values after the initiation of symptomatic treatment to 'missing' and using a mixed model for repeated measures. Here, all values for $Y_t$ after symptomatic medication was initiated are set to 'missing'. Thereafter, a mixed model for repeated measures with $Y_t$ as longitudinal outcome and treatment assignment $Z$, timepoints $t$ and the interaction between $Z$ and $t$ as predictors with an unstructured correlation matrix over time and allowing different variances at each timepoint.

## 2.4. Simulation study

To investigate the performance of the above-described candidate estimators, we implemented a simulation study based on the DAGs described above. The data generating model is similar to a simulation study published by Lasch et al [10] using the CDR-SB as the endpoint of interest but includes minor modifications to fit the ADAS-Cog 13 as the endpoint of interest, and to allow initiation of symptomatic medications at multiple timepoints. Therefore, longitudinal patient-level data on the ADAS-Cog 13 scale were simulated based on a non-linear mixed effects model for a randomized parallel group trial investigating a test drug in comparison to placebo. The data-generating model is described in detail in the following sections.

### 2.4.1. Baseline

While clinical trials in the prodromal AD population generally do not use ADAS-Cog 13 as an inclusion criterion, the patient characteristics at baseline were simulated in line with trials [22, 23] and cohorts [24] recruiting a comparable population.

For each patient $i \in \{1, \ldots, n\}$, the treatment allocation $treat \in \{1,0\}$ (placebo or active treatment) was simulated as Bernoulli distributed variables with fixed probabilities of 0.5. The baseline ADAS Cog 13 value $Y^*_{0,i}$ (where $\square^*$ indicates for all values that they are the counterfactual values unaffected by symptomatic treatment initiation), and a natural decline rate $\alpha_i$ were simulated as correlated random intercept and slope from a multivariate normal distribution.

### 2.4.2. Disease Progression

The natural course of Alzheimer's Disease has been studied extensively, using data from longitudinal cohorts and control arms of clinical trials, and detailed disease progression models have been developed. For this article, we base the choice of our disease-progression model on the work of (i) Rogers et al. [25] on combining patient-level and summary data in a meta-analysis for Alzheimer's disease modelling and simulation and (ii) Samtami et al on a model for disease progression in patients from the Alzheimer's Disease Neuroimaging Initiative (ADNI) study. Both articles use the ADAS-Cog scale as the endpoint of interest.

Correspondingly, for simulating the longitudinal ADAS-Cog 13 scores, a beta regression model with Richard's logistic function as link function was used as the underlying disease progression model. In short, the ADAS-Cog 13 value $Y^*_{t,i}$ for each patient is generated from the beta distribution, which depends on the mean $\bar{Y}^*_{t,i}$. This mean, in turn, is transformed by the link function (Richard's function) and modelled based on the natural decline rate and the treatment effect as shown in the following equations. According to our motivating example, the treatment is disease modifying and thus influences the natural decline rate of the respective patient.

For time points $t$ (years) $\in T = \{0, 0.25, 0.5, 0.75, 1, 1.25, 1.5, 1.75, 2\}$ longitudinal data for ADAS-Cog 13 are simulated, but only data at $t \in \{0, 0.5, 1, 1.5, 2\}$ are considered to observed and used in the analysis. Following the notation used by Rogers et al., for simulating the underlying ADAS-Cog 13 scores $Y_{t,i}^*$, for each patient $i$ at time $t$, the normalized variable $Y_{t,i}^*/85$ is simulated using a beta-regression model

$$\frac{Y_{t,i}^*}{85} \sim Beta\left(\frac{\bar{Y}_{t,i}^*}{85} * \tau, \left(1 - \frac{\bar{Y}_{t,i}^*}{85}\right) * \tau\right)$$

where $\frac{\bar{Y}_{t,i}^*}{85} := E\left[\frac{Y_{t,i}^*}{85}\right]$ is the mean of $Y_{t,i}^*$ and the beta-distribution Beta(a, b) is parameterized with $a = \frac{\bar{Y}_{t,i}^*}{85} * \tau$ and $b = (1 - \frac{\bar{Y}_{t,i}^*}{85}) * \tau$ so that the variance for patient $i$ at time $t$ is

$$V\left[\frac{Y_{t,i}^*}{85}\right] = \frac{\frac{\bar{Y}_{t,i}^*}{85} * \left(1 - \frac{\bar{Y}_{t,i}^*}{85}\right)}{\tau + 1}$$

where $\tau$ is a nuisance parameter for scaling the variance. For modelling the conditional expectation for patient $i$, $\bar{Y}_{t,i}^*$, Richard's logistic model

$$g(x) = \log\left[\left(\frac{x^\beta}{(1-x^\beta)}\right)^{1/\beta}\right]$$

was used as a link function to ensure a sigmoidal shape of the ADAS-Cog 13 progression and account for the boundedness of the ADAS-COG 13 scale at 85 points.

For a timepoint $t \in T$ with $t > 0$, we model the link-transformed expectation $g(\bar{Y}_{t,i}^*)$ based on the observed value $Y_{t_{prev},i}^*$ at the previous timepoint $t_{prev} < t$ and the decline rate of the patient as

$$g(\bar{Y}_{t,i}^*) = intercept_{i,t} + \alpha_i * \frac{t - t_{prev}}{52} * (E_{DM})^{treat_i}$$

Particularly, the intercept is calculated from the underlying ADAS-Cog 13 score at the previous timepoint $t_{prev}$ as

$$intercept_{i,t} = g\left(Y^*_{t_{prev},i}\right) = \log\left[\left(\frac{\left(\frac{Y^*_{0,i}}{85}\right)^{\beta}}{1-\left(\frac{Y^*_{0,i}}{85}\right)^{\beta}}\right)^{\frac{1}{\beta}}\right]$$

Under the Null hypothesis, the treatment effect is set to $E_{DM} = 1$ and under the alternative hypothesis, the treatment effect will be set to $E_{DM} = 0.5$, which corresponds to slowing down the decline rate by 50%.

### 2.4.3. Simulation parameters

Based on published clinical data [3, 26, 27], in our simulation study we aim for a baseline mean of 27, a decline of approximately 6 points in the ADAS-Cog 13 after two years compared to baseline and a standard deviation of the ADAS-Cog 13 value after 2 years of approximately 14 in the placebo group. To achieve these characteristics, we set the mean for the baseline severity to 27 and the mean decline rate for each patient to 0.23, representing an expected decline (that is, increase in ADAS-Cog 13) of approximately 6 points over the two-year trial duration. The covariance matrix for the correlated random effect of baseline value and decline rate was set similar to the work of Conrado et al. [28] that focused on CDR-SB as the endpoint of interest. The variance of the decline rate was chosen based on the signal to noise ratio for the decline rate for the CDS-SB as an endpoint. Here, Conrado et al [28] calculated a mean decline rate for the beta-regression model of 0.2 with a variance of 0.062, resulting in a mean/variance ratio of 3.2. Consequently, setting the mean of the decline to 0.23, we assume a variance of 0.072 to achieve the same signal to noise ratio. In line with Conrado's work, we chose a covariance between baseline ADAS-Cog 13 and decline rate of 0.69 to ensure a correlation of 0.37, resulting in a covariance matrix of $\begin{pmatrix} 49 & 0.69 \\ 0.69 & 0.072 \end{pmatrix}$. For the baseline severity, the underlying multivariate distribution was truncated to ensure that the baseline severity was between 10 and 50, reflecting a likely range reflecting the inclusion criteria for a trial in early AD.

For the disease progression model, the parameter $\beta$ from Richard's link function and the nuisance parameter $\tau$ from the beta-distribution that scales the variance need to be specified.

In line with the article from Samtami et al that discussed the inflection points of the link function, $\beta = 2.4$ was chosen as follows. Samtami et al cite a range from 40 to 42 as inflections points for the ADAS-Cog 11, which is bounded at 70. This corresponds to an inflection point of $\frac{42}{70} = 0.6$ for the normalized values of the ADAS-Cog 11. Correspondingly, we assume the same relative inflection point for the normalized values (divided by 85) of the ADAS-Cog 13 scale of $0.6$. Modifying the formula for the inflection points provided in table 1 of Samtami et al, this corresponds to an inflection point of $0.6 = (\frac{85^\beta}{1+\beta})^{\frac{1}{\beta}}$, from which $\beta = 2.4$ follows.

To approximate an appropriate value for $\tau$, we used the data from the placebo group at baseline and after 104 weeks of treatment from the publications of Wessels et al[26], summarizing trials in mild to moderate Alzheimer's disease. In the Amaranth trial published by Wessels et al, a standard deviation for the ADAS-Cog 13 of approximately 14 was seen for the measurements after two years, which we chose as the target for the variability of $Y^*_{2,i}$. Given that the variability of the decline rate contributes to the variability of $Y^*_{2,i}$, we calculated $\tau = 174.15$ as nuisance parameter in the beta-distribution to achieve the desired variability in $Y^*_{2,i}$.

### 2.4.4. Intercurrent event – start of symptomatic treatment

The probability to initiate symptomatic treatment increases with disease severity and is subject to variability. We defined the average effect of symptomatic treatments on the ADAS-Cog 13 as 2.6, based on the between-arm difference in large meta-analyses [29, 30]. The results of same meta-analysis allow us to apply this effect across the spectrum of severity we investigate (i.e. without interaction with the underlying severity value).

To calculate the observed ADAS-Cog 13 values for each patient and time point, the intercurrent event of initiating symptomatic treatment needs to be incorporated. At the observed timepoints $T = \{0, 0.5, 1, 1.5\}$, the initiation for each patient was modelled as a Bernoulli random variable $Sym_{t,i} \sim Ber(p_{t,i})$ with success probability $p_{t,i}$ based on a sigmoid function with inflection point at 29 using the underlying ADAS-Cog 13 score $Y^*_{t,i}$ at the given time point as:

$$p_{t,i} = \frac{1}{1 + \exp\left(-(Y^*_{t,i} - 29)\right)}$$

which is between 0 and 1. It is worth noting that the probability is inversely correlated with the ADAS-Cog 13 score at the respective time point (i.e. the worse a patient performs, the most likely symptomatic treatment is started). In the presence of a disease-modifying treatment effect, this implies that patients assigned to the treatment group are less likely to start symptomatic treatment as compared to the placebo group. In our simulation, patients stay on symptomatic treatment once they have started, so from $Sym_{t,i} = 1$ for a given $t \in T$ it follows that $Sym_{\tilde{t},i} = 1$ for all $\tilde{t} \in T$ with $\tilde{t} > t$.

For each patient $i$, a random effect of the symptomatic treatment $E_{sym,i} \sim N(-2, (0.1)^2)_{[-4,0]}$ is simulated from a truncated normal distribution between -4.6 and 0 with mean -2.6 in line with the published magnitude of the symptomatic effect and standard deviation 2. For all time points $t$, where symptomatic treatment is taken ($Sym_{t,i} = 1$), this effect is added to the underlying ADAS-Cog 13 score $Y^*_{t,i}$ to generate the observed ADAS-Cog 13 score $Y^*_{t,i}$, without modifying the decline parameter:

$$Y_{t,i} = Y^*_{t,i} + Sym_{t,i} * E_{sym,i}$$

Note, that (i) within each patient the exact same effect is used for all timepoints after the initiation of symptomatic treatment and (ii) this assumes that the effect of the symptomatic treatment is equal at all time points in expectation since the distribution of $E_{sym,i} \sim N(-2.6, (3)^2)_{[-8.6, 5.4]}$ is not time-dependent. In particular, the effect of the symptomatic treatment is assumed to be equal regardless of the symptomatic treatment is initiated at $t = 0.5$, or $t = 1.5$. The treatment assignment is not used in the calculation of the effect of symptomatic treatment, in line with the assumption that no pharmacological interaction exists between the two. Subsequently, the observed values are truncated at 0 and 85, and both $Y_{t,i}$ and $Y^*_{t,i}$ are rounded to fit the ADAS-Cog 13 scale (0, 1, 2, … 85). For the comparison of the estimators, only the baseline values, the occurrence of the intercurrent event and the observed ADAS-Cog 13 values, $Y_{t,i}$, at the modelled time points will be used, whereas the true value of the estimand that employs the hypothetical strategy will be calculated using the underlying values $Y^*_{t,i}$.

### 2.4.5. Simulation scenarios

To explore the performance of the above-described estimators and compare the proposed extensions utilizing the fact that the same symptomatic treatment can be initiated at several timepoints (in contrast to assuming that different mediators occur at different time points) with the established approach, we investigate a clinically realistic scenario with simulation parameters chosen based on published, comparable trials in Alzheimer's Disease, both in presence (treatment effect $E_{DM} = \frac{1}{2}$ reducing the natural decline to one-half in the treatment arm, see equation (6)) and in absence of a treatment effect (treatment effect $E_{DM} = 1$). The sample size is chosen to be 154 to provide a nominal power of 90% under the alternative hypothesis of a true disease-modifying effect.

To investigate the generalizability of our findings to different disease settings where a hypothetical strategy for handling some intercurrents may be of primary or secondary interest, we explored the performance of the estimators in a range of scenarios varying (i) the standard deviation of $Y_2$, (ii) the proportion of variability that is contributed by the "clinical variability" induced by the variability in the decline rate (scaled by the standard deviation of the decline rate at baseline), as compared to the variability induced by the beta-distribution (scaled by the nuisance parameter τ), and (iii) the variability of the effect of the symptomatic medication.

For each of the modifications, heatmaps were created comparing their empirical standard error to (i) the established g-estimation approach for longitudinal de-mediation and (ii) a linear model based on the unaffected values $Y_2^*$. Each cell of the heatmaps corresponds to a set of 10000 trials. The sets differ in two dimensions. The first is the standard deviation of the effect of starting a symptomatic treatment (with a standard deviation from 1 to 15) and the second is different in the two series of heatmaps:

(a) fixing the overall variability of the unaffected $Y_2^*$, at 14, the proportion of such variability that is due to variability in decline rate is varied between 50% and 95%;

(b) fixing the proportion of variability that is due to variability in decline rate, higher variabilities of the unaffected $Y_2^*$ are investigated (standard deviations of 10 and 13.5).

### 2.4.6. Expected value

The expected value of the estimand θ is 0 under the Null hypothesis. Under the alternative hypothesis, the expected value of the estimand is not a parameter of the data-generating mechanism, as the disease-modifying treatment effect is modelled as a multiplicative effect on the decline in the beta-regression model (see section on the 'disease progression'). To derive the expected value for each simulation scenario, we estimated the true value based on a trial with 100,000,000 patients using a linear regression model for the underlying values of the outcome before applying the effect of the symptomatic medication, $Y_2^*$, as dependent variable and treatment and baseline severity as linear predictors.

## *2.5. Performance criteria*

To compare the candidate estimators for the estimand of interest, we primarily investigate the following properties that are averages (or functions of averages) over all simulation runs for the same scenario [31]:

- bias, expressed in ADAS-Cog 13 and estimated as $\frac{1}{nsim}\sum \hat{\theta}_k - \theta$ (where $nsim$ is the number of simulations, $\hat{\theta}_k$ is the value of the estimand estimated by each method in a specific simulation run $k$ and θ is the true value of the estimand in the respective simulation scenario);

- empirical standard deviation of the estimates $\hat{\theta}_k$, and the averages of the model-derived, jackknife-derived and boostrap-derived standard error estimates $\frac{1}{nsim}\sum \widehat{se(\theta)}_k$

- empirical type I error / empirical power, measured as the proportion of cases where the null hypothesis was rejected. This will be reported and interpreted as empirical type I error rate for the scenarios under the null hypothesis and as empirical power under the alternative hypothesis.

For all performance criteria, Monte Carlo estimates of the Standard Errors have been computed [31]. Additionally, coverage and confidence-interval length have been visualized with zip plots [31].

## 3. Results

In the following, we first discuss the results for the clinically realistic scenario based on the motivating example in Alzheimer's disease, before we explore the performance of the estimators under different assumptions for the variability and sources of variability.

### 3.1. Clinically realistic scenario

#### 3.1.1. Null hypothesis

Under the null hypothesis in the realistic scenario, the true value of the estimand is 0. The performance measures for the estimators tested are reported in Table 1. All g-estimation methods preserve the one-sided type I error rate at 0.025 if the standard error for the effect estimate is estimated using jackknife or bootstrapping. All estimators show a negligible bias on the ADAS-Cog 13 scale used as the primary endpoint.

*Table 2: Simulation results under the null hypothesis*

|  | Empirical Bias* | Empirical standard deviation of the effect estimate | Mean model derived Standard Error | Mean Bootstrap based Standard Error | Empirical Type I Error based on bootstrapping | Mean Jackknife based Standard Error | Empirical Type I Error based on jackknife |
|---|---|---|---|---|---|---|---|
| MMRM | -0.0185 | 2.747 | 2.722 | NA | 0.0259[#] | NA | NA |
| Established g-estimation approach for longitudinal de-mediation | 0.0187 | 2.145 | 2.105 | 2.169 | 0.0217 | 2.207 | 0.0188 |
| Modification 1 Averaging the de-mediation on $Y_t$ | 0.0175 | 2.112 | 2.088 | 2.087 | 0.0249 | 2.108 | 0.0238 |
| Modification 2 Averaging the de-mediation effects on $Y_2$ | 0.0174 | 2.124 | 2.092 | 2.108 | 0.0242 | 2.135 | 0.0216 |
| Modification 3 Iterative averaging the de-mediation effects on $Y_2$ | 0.0170 | 2.1229 | 2.091 | 2.105 | 0.0245 | 2.132 | 0.0221 |

|  | Empirical Bias* | Empirical standard deviation of the effect estimate | Mean model derived Standard Error | Mean Bootstrap based Standard Error | Empirical Type I Error based on bootstrapping | Mean Jackknife based Standard Error | Empirical Type I Error based on jackknife |
|---|---|---|---|---|---|---|---|

* The expected value of the effect (active treatment – placebo) under the alternative hypothesis is 0 on the ADAS-Cog 13 scale, where lower values correspond to a less sever disease. The bias is calculated as model estimate – expected value.  A positive bias corresponds to an underestimation of the effect.

#: model-based empirical type I error is reported for the observed values estimator and the MMRM as the model-derived standard errors are unbiased.

### 3.1.2. Alternative hypothesis

Under the alternative hypothesis in the realistic scenario, the true value of the estimand is -5.83. As shown in Table 2., the reference method that is commonly used for analysing these types of trials, setting values after the initiation of symptomatic treatment to 'missing' and using a mixed model for repeated measures, underestimates the treatment effect approximately 15%, which is in line with previous findings in a simpler setting [10]. In contrast, all de-mediation methods display a neglectable bias, with a slightly higher bias of the established approach. Regarding the precision of the estimates, the MMRM a markedly higher empirical standard deviation of the estimate as compared to the de-mediation approaches. The modifications of the established g-estimation approach for longitudinal de-mediation show slightly different performances. As expected, the model-based standard error estimates underestimate the true standard deviation of the effect estimates, since they do not take into account the uncertainty of the de-mediation step. In line with this, all de-mediation methods have under-coverage when using the model-based SE (Figure 2). In contrast, the standard error estimates derived via the jackknife approach overestimate the true standard deviations of the effect estimates. Only the standard error estimates from bootstrapping are unbiased. Correspondingly, comparing the power of the investigated methods based on the bootstrap-based standard errors, shows that the established g-estimation approach for longitudinal de-mediation loses approx. five percentage points of power as compared to the nominal power of 90%. Here, all proposed modifications recover the power loss partly (averaging the de-mediation effect on Y2, either in one run or iteratively) or in full (averaging the de-mediation effects at the respective timepoints). As shown in Figure 2, the first modification – averaging the most proximal effect – has ideal coverage both with SE estimated through jackknife and bootstrap.

*Figure 2: Zip plot for the simulations under the alternative hypothesis. The CI of coverage are in green for optimal coverage, in red for undercoverage and in blue for overcoverage.*

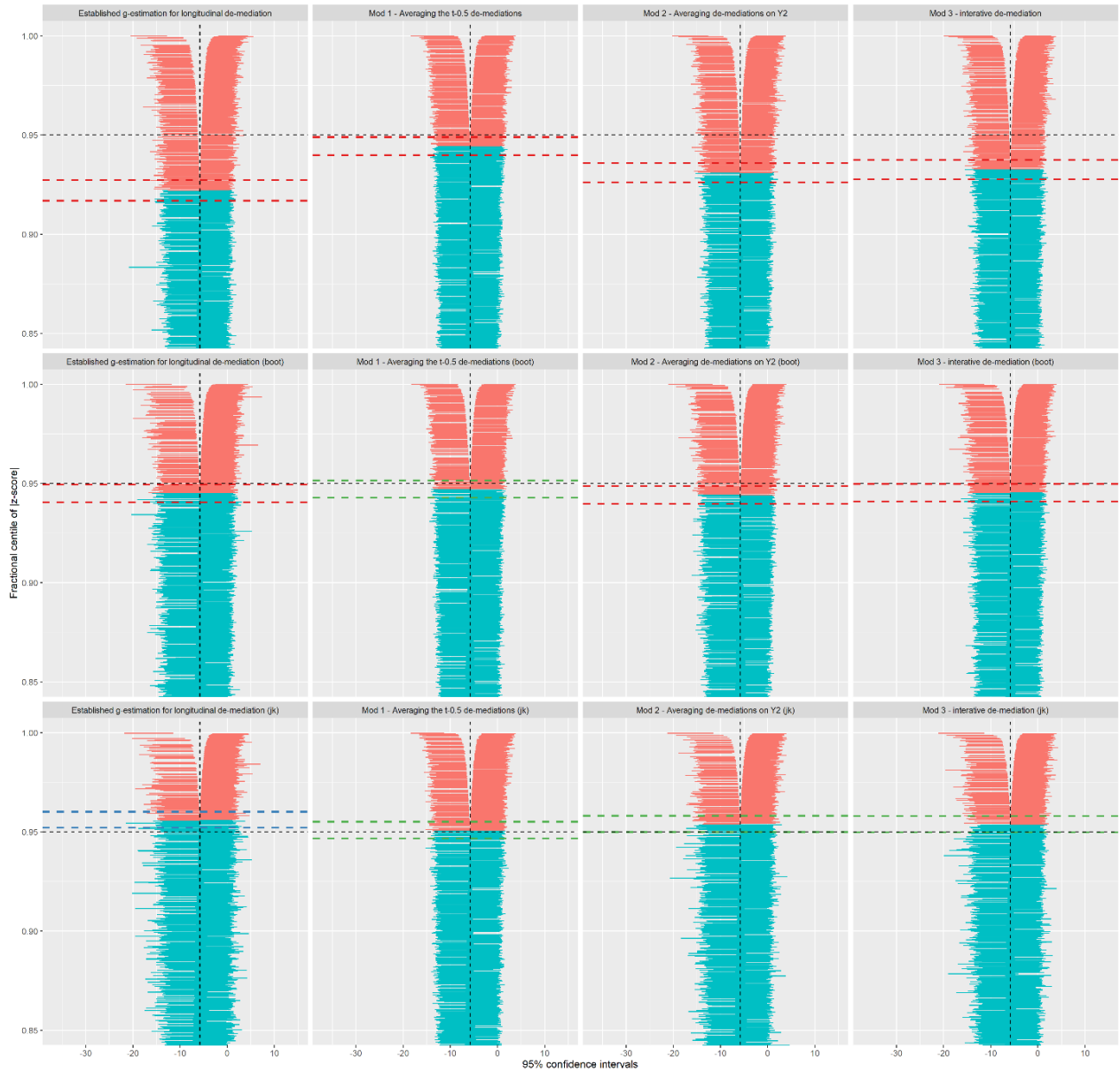

*Table 3: Simulation results under the alternative hypothesis*

|  | **Empirical Bias*** | **Empirical standard deviation of the effect estimate** | **Mean model derived Standard Error** | **Mean Bootstrap based Standard Error** | **Empirical Power based on bootstrapping** | **Mean Jackknife based Standard Error** | **Empirical Power based on jackknife** |
|---|---|---|---|---|---|---|---|
| MMRM | 0.858 | 2.301 | 2.275 | NA | 0.58# | NA | NA |
| Established g-estimation approach for longitudinal de-mediation | 0.115 | 1.988 | 1.818 | 1.981 | 0.8523 | 2.146 | 0.7973 |
| Modification 1 Averaging the de-mediation on $Y_t$ | 0.052 | 1.825 | 1.800 | 1.818 | 0.8929 | 1.859 | 0.8871 |
| Modification 2 Averaging the | 0.059 | 1.912 | 1.806 | 1.888 | 0.8851 | 2.003 | 0.8538 |

|  | Empirical Bias* | Empirical standard deviation of the effect estimate | Mean model derived Standard Error | Mean Bootstrap based Standard Error | Empirical Power based on bootstrapping | Mean Jackknife based Standard Error | Empirical Power based on jackknife |
|---|---|---|---|---|---|---|---|
| de-mediation effects on $Y_2$ | | | | | | | |
| Modification 3 Iterative averaging the de-mediation effects on $Y_2$ | 0.056 | 1.901 | 1.805 | 1.882 | 0.8859 | 1.989 | 0.8585 |

* The expected value of the effect (active treatment – placebo) under the alternative hypothesis is -5.83 on the ADAS-Cog 13 scale, where lower values correspond to a less sever disease. The bias is calculated as model estimate – expected value. A positive bias corresponds to an underestimation of the effect.

#: model-based power is reported for the observed values estimator and the MMRM as the model-derived standard errors are unbiased.

### 3.2. Exploring the impact of variability

The exploration across different levels and sources of variability (see heatmaps in the Appendix) generally confirms that the modifications increase precision over the established g-estimation approach for longitudinal de-mediation. The gain in precision of modification 1 over the established g-estimation approach for longitudinal de-mediation is confirmed consistently whereas, for modifications 2 and 3, the magnitude of the advantage compared to the established approach depends on the underlying variability of $Y_2$.

The comparison of the modifications with the linear model based on the unaffected values $Y_2^*$ shows a more complex picture. While for lower variabilities all modifications and the established g-estimation approach for longitudinal de-mediation show a small reduction in precision (as expected), for higher variabilities they show a slight advantage in precision. This decrease in empirical standard error is accompanied with an increase in bias of all methods for higher variabilities. This underestimation of the true treatment effect leads to a smaller empirical standard deviation of the estimates.

As an additional analysis of the robustness of our findings, we explored a scenario where the symptomatic medication was initiated not using on a probabilistic model based on Y, but a deterministic model with a threshold of 40.5 - all patients with an observed value of Y higher at 40.5 would initiate the symptomatic medication at the respective timepoint (see supplementary material). These results indicate a very small increase in the bias on the absolute scale under the most extreme mechanism for initiation of alternative treatments, but the bias that does not favour active treatment.

As such, this provides additional reassurance that these methods could be suitable for regulatory decision making.

## 4. Discussion

In clinical trials in Alzheimer's Disease, the initiation of symptomatic medication is an important intercurrent event that can occur at different timepoints during the clinical trial, for which a hypothetical strategy is of regulatory interest. In trials for early AD, it is likely that the effect of the symptomatic medication on the outcome is the same regardless of the timepoint of initiation. Building on previous work that has shown that longitudinal de-mediation approaches are a promising estimator for the treatment effect of interest in these setting, in this article we explore extensions to the established g-estimation approach for longitudinal de-mediation, utilising the assumption that the effect of the mediator is the same at all possible initiation timepoints. The key extension as compared to established approaches is to use information from all timepoints simultaneously by averaging the estimated mediator effects before de-mediating the outcome. Here, we explored three modifications. Modification 1 estimates the mediator effect directly on outcome at the subsequent timepoint, while modification 2 estimates the mediator effect on the outcome of the last timepoint of interest, before the estimated effects are averaged. Modification 3 explores an iterative optimisation of the estimation of the mediator effects of modification 2 by using the information on the mediator effects at different timepoints also for the estimation of the mediator effects.

Our simulation study in the clinically realistic setting has confirmed findings from previous work in a simpler setting that standard methods usually applied to these types of trials (setting values after initiating symptomatic treatment to 'missing' and applying a MMRM) underestimate the treatment effect by a relevant amount and lead to power loss under the alternative hypothesis. In contrast, our simulation study has shown that none of the estimators tested has a relevant bias. For estimating the standard errors, re-sampling-based methods are necessary, as the model-based standard errors underestimate the variability. While bootstrap and jackknife by and large accurately estimate the standard error, we noted a slight underestimation by bootstrap and overestimation by the jackknife approach. Using re-sampling-based standard errors, all modifications increase the precision of the effect estimate as compared to the established g-estimation approach for longitudinal de-mediation that assumes different effects of the mediator at each timepoint, and correspondingly show a power advantage. In our setting, modification 1 that estimates the mediator effect directly on outcome at the subsequent timepoint, shows the best performance. This is likely due to the fact that the mediator effects can be more precisely estimated by using the subsequent timepoint (instead of the last timepoint of interest), as these carry less noise.

This gain in precision of modification 1 over the established g-estimation approach for longitudinal de-mediation is confirmed consistently by the exploration for higher variabilities and different sources of the variability. In contrast, for modifications 2 and 3, the magnitude of the advantage compared to the established approach depends slightly on the underlying variability of $Y_2$ and increases with increasing variability.

A critical assumption of the investigated modifications is the constancy of the mediator effect at the different timepoints. This assumption might be violated in case the severity is a predictor for the effect of the mediator, e.g. in a rapidly progressive disease where the symptomatic medication does not work equally in disease stages that are transitioned through within the duration of a trial.

In particular, we have assumed that the onset of the full effect of the mediator happens before the next consecutive timepoint and that there is no reduction in effect over time. Depending on the duration of the trial, frequency of measurement timepoints, the disease and the symptomatic medication of interest, these assumptions might not hold true. Further research is needed to explore performance of the proposed estimators in these settings.

A second question for further investigation is the performance of the model for smaller sample sizes and / or a higher number of covariates that would need to be included in the mediation models. If only a small number of patients initiates symptomatic medication at a given timepoint, and the assumed causal structure is more complex, justifying inclusion of additional covariates in the model (sex, geographic region, severity on other symptomatic dimensions, etc.), model performance might degrade.

A third direction for further research is the inclusion of prior information about the mediator effect in the de-mediation models. In our motivating example, the symptomatic medication is from a class of approved medicines, and hence a notable amount of clinical evidence can be assumed to be present about its efficacy. Incorporating this evidence could further improve the precision gains. Understanding how to best incorporate external evidence, possibly using Bayesian methods, and how this impacts the performance under different scenarios is a promising way for extending de-mediation approaches further. Generally, the possibility to explicitly model (and evaluate) the effect of the mediator on the outcome, and to incorporate external evidence, could be an additional advantage of de-mediation approaches for estimating a hypothetical strategy as compared to missing data techniques that deserves further attention.

# 6. Supplementary material

*6.1. Varying SD of the effect of symptomatic treatment and SD of $Y_2^{\square}$:*

### 6.1.1. Standard deviation compared to established g-estimation method for longitudinal de-mediation

**Figure S1.** Difference (in % of the SD of the estimates of the established g-estimation method for longitudinal measurements) between the SD of the estimates of the averaging method and the SD of the estimates of the established g-estimation method for longitudinal measurements, in sets of 10.000 simulated trials per combination of parameters.

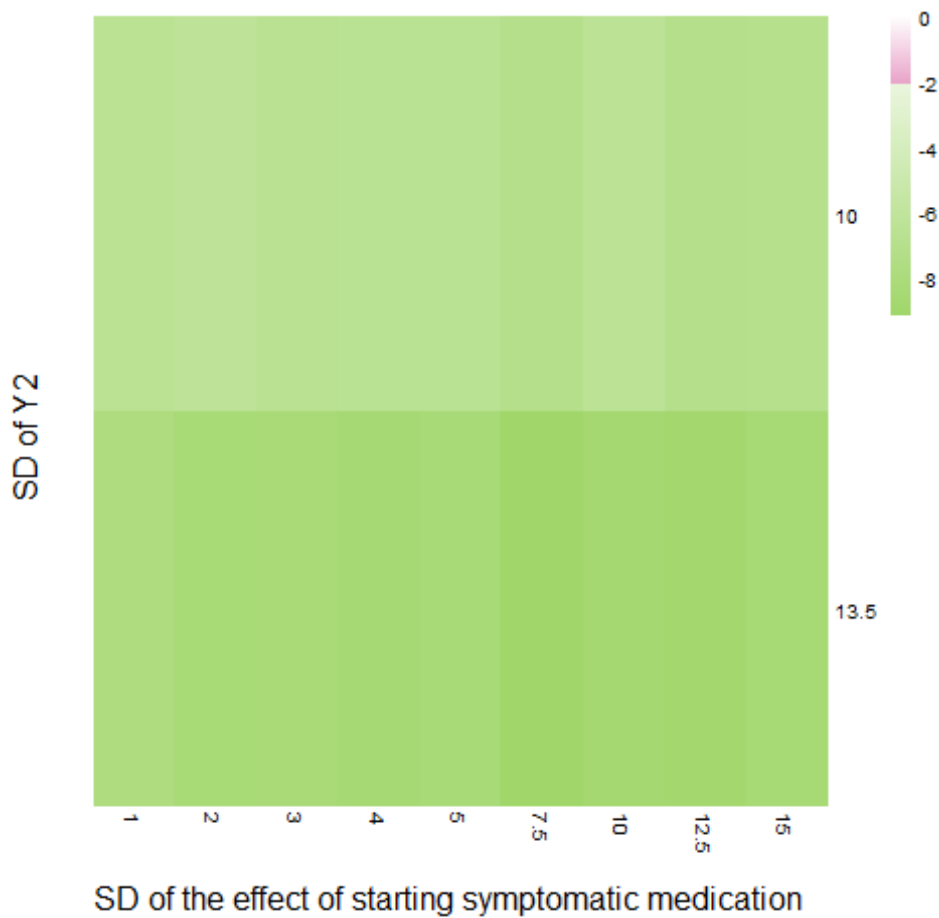

**Figure S2.** Difference (in % of the SD of the estimates of the established g-estimation method for longitudinal measurements) between the SD of the estimates of the averaging $Y_2^{\square}$ method and the SD of the estimates of the established g-estimation method for longitudinal measurements, in sets of 10.000 simulated trials per combination of parameters.

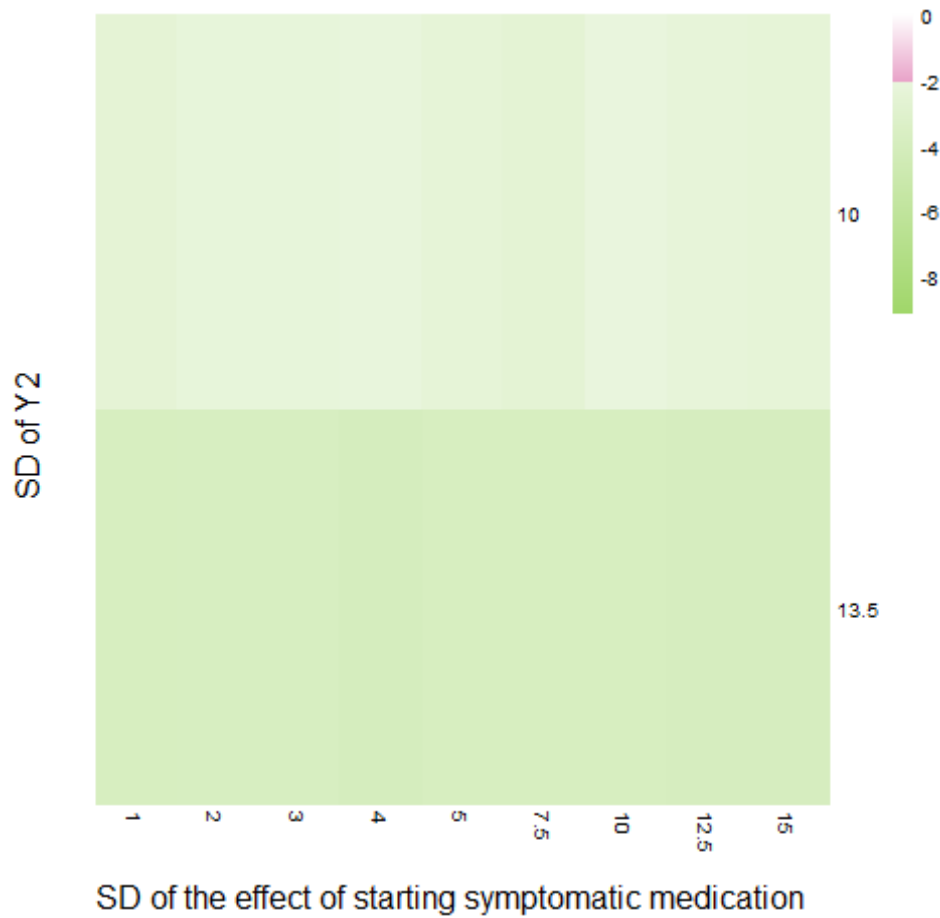

**Figure S3.** Difference (in % of the SD of the estimates of the established g-estimation method for longitudinal measurements) between the SD of the estimates of the iterative method and the SD of the estimates of the established g-estimation method for longitudinal measurements, in sets of 10.000 simulated trials per combination of parameters.

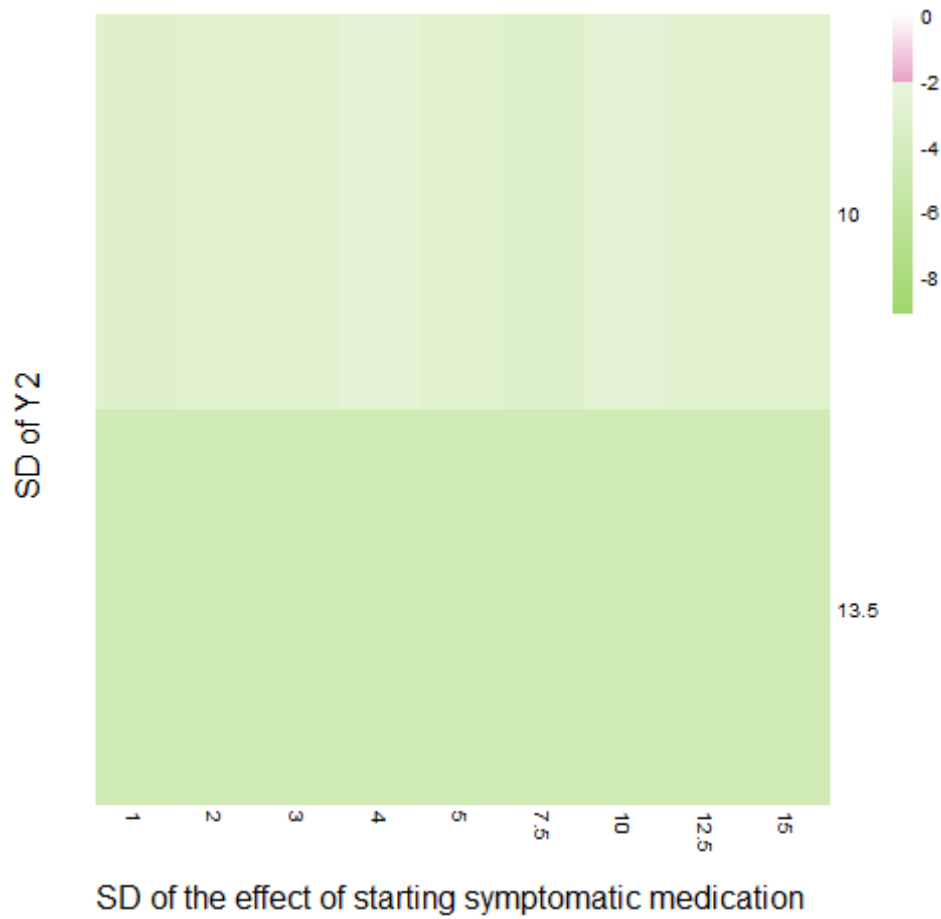

### 6.1.2. Standard deviation compared to a linear model based on the unaffected values $Y_2^*$

**Figure S4.** Difference (in % of the SD of the estimates of the linear model based on the unaffected values $Y_2^*$) between the SD of the estimates of the averaging method and the SD of the estimates of the linear model based on the unaffected values $Y_2^*$, in sets of 10.000 simulated trials per combination of parameters.

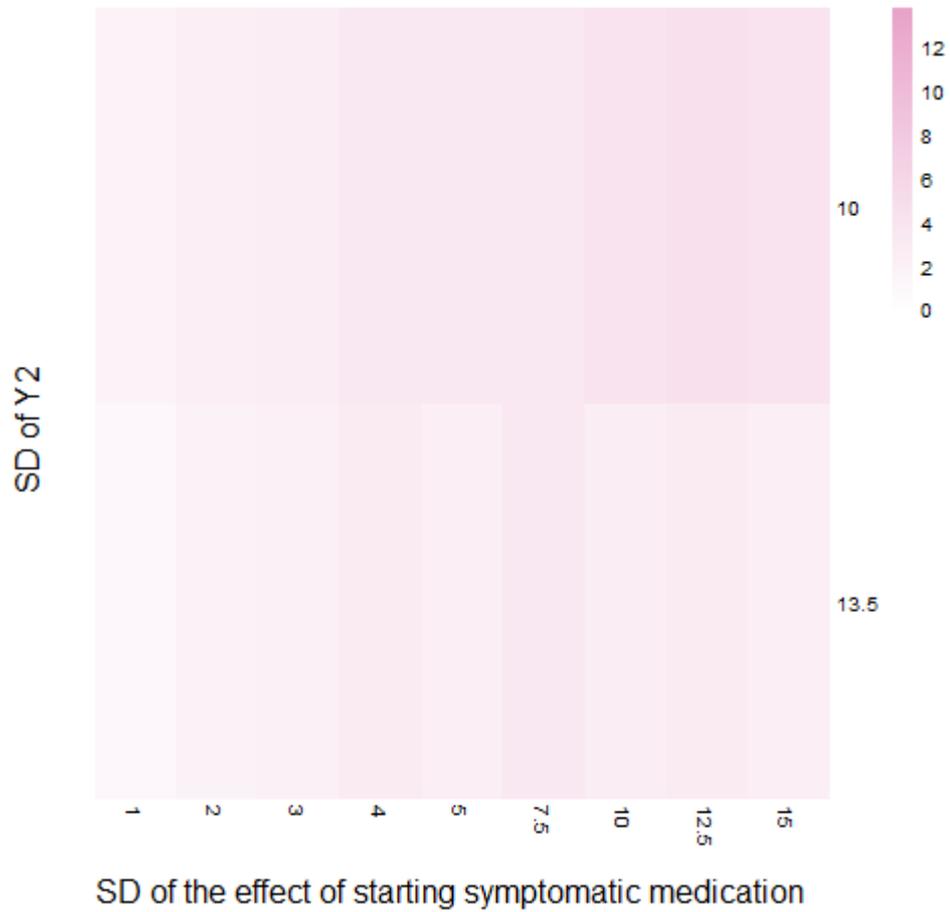

**Figure S5.** Difference (in % of the SD of the estimates of the linear model based on the unaffected values $Y_2^*$) between the SD of the estimates of the averaging $Y_2^{\square}$ method and the SD of the estimates of the linear model based on the unaffected values $Y_2^*$, in sets of 10.000 simulated trials per combination of parameters.

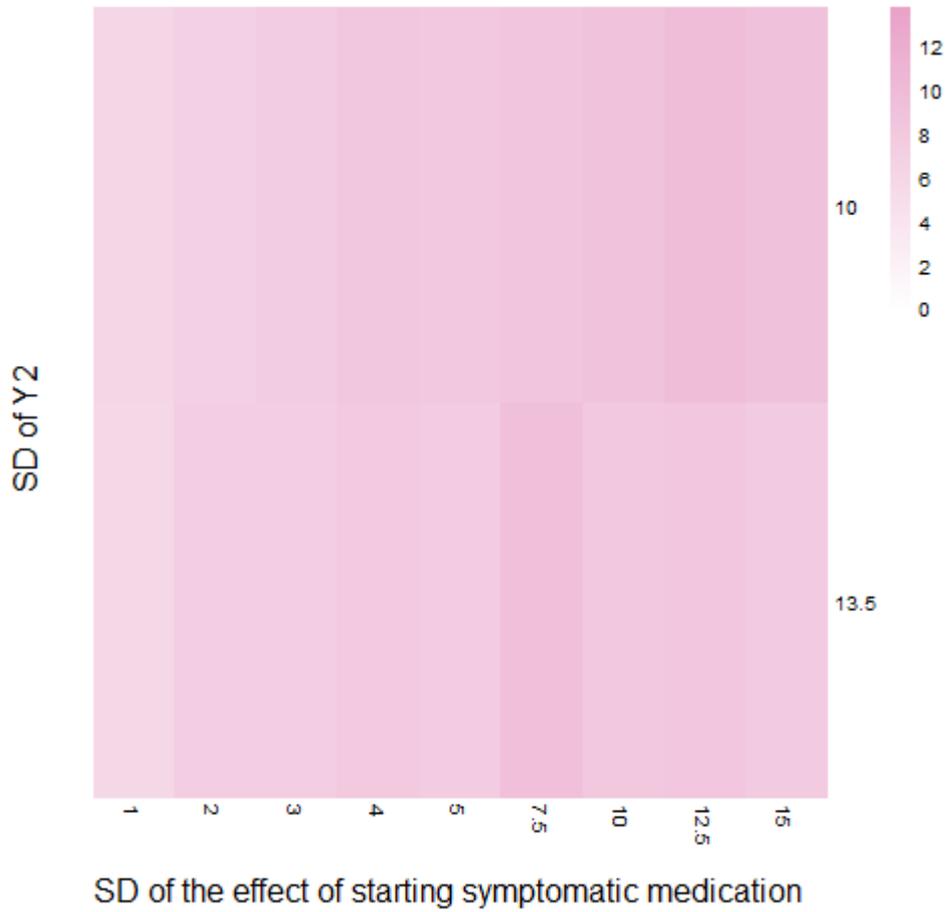

**Figure S6.** Difference (in % of the SD of the estimates of the linear model based on the unaffected values $Y_2^*$) between the SD of the estimates of the iterative method and the SD of the estimates of the linear model based on the unaffected values $Y_2^*$, in sets of 10.000 simulated trials per combination of parameters. The X axis labels indicate SD of the symptomatic effect, Y-axis labels indicate the SD of $Y_2^{\square}$.

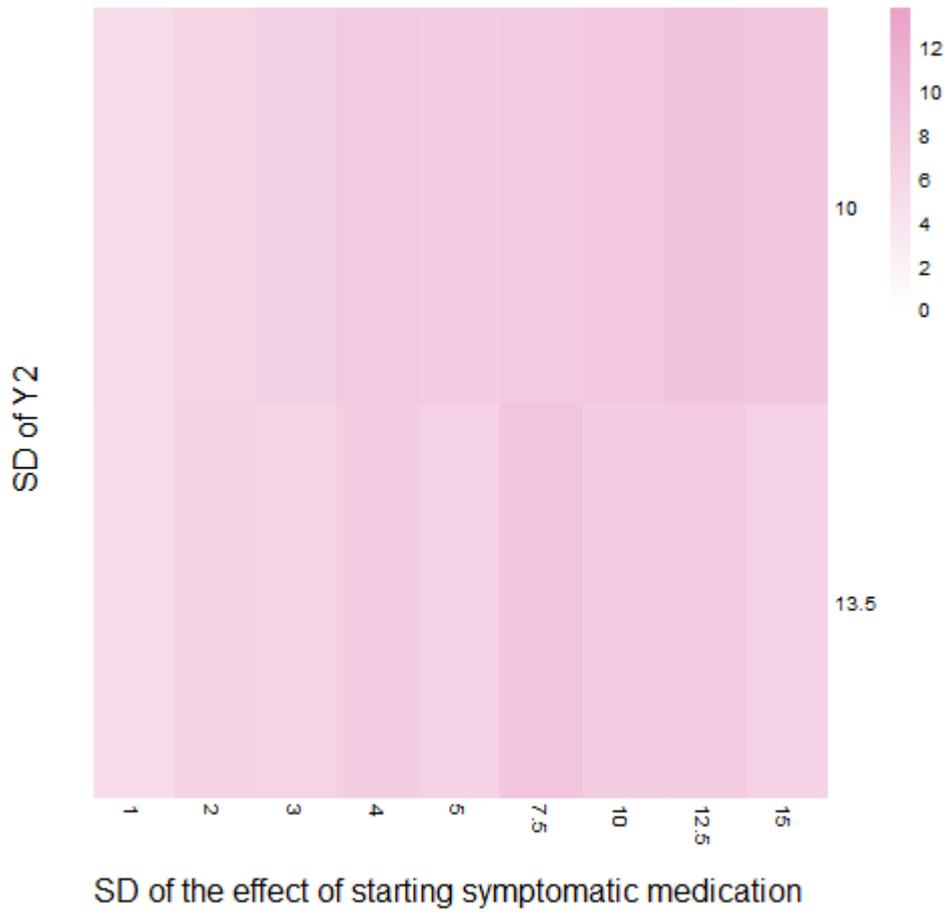

**Figure S7.** Difference (in % of the SD of the estimates of the linear model based on the unaffected values $Y_2^*$) between the SD of the estimates of the established g-estimation method for longitudinal measurements and the SD of the estimates of the linear model based on the unaffected values $Y_2^*$, in sets of 10.000 simulated trials per combination of parameters. The X axis labels indicate SD of the symptomatic effect, Y axis labels indicate the SD of $Y_2^{\square}$.

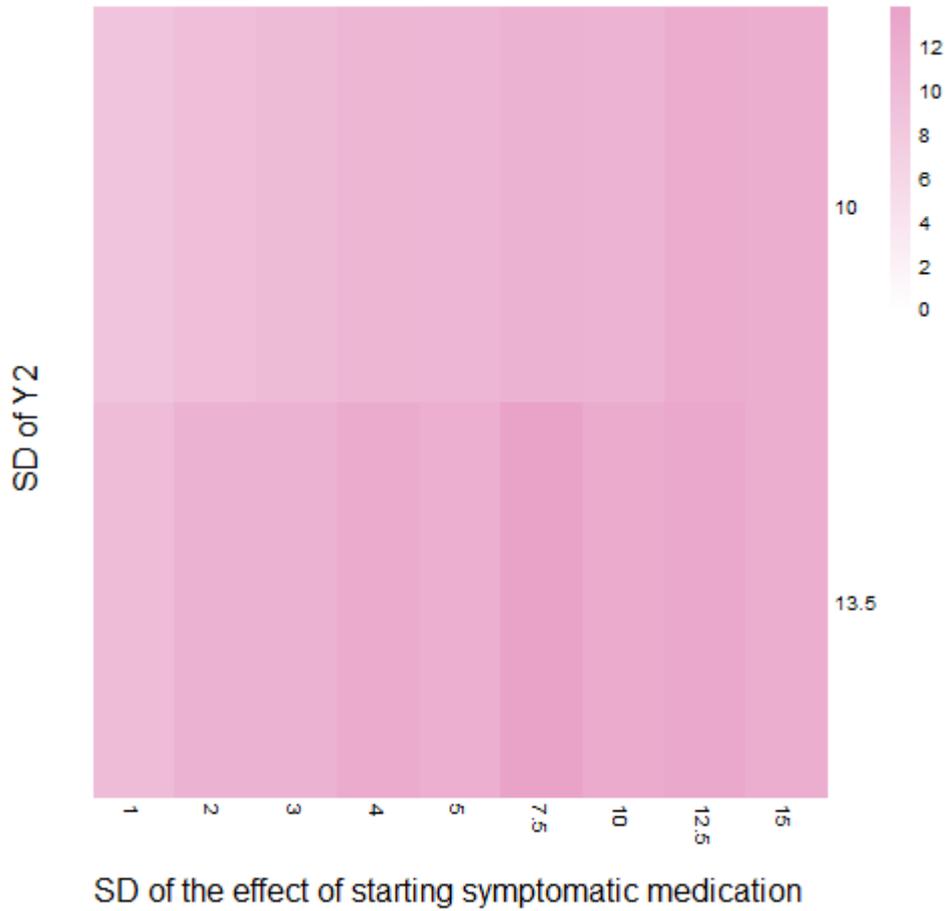

## 6.2. Varying SD of the effect of symptomatic treatment and the proportion of SD of $Y_2^{\square}$ due to heterogeneity in natural decline

### 6.2.1. Standard deviation compared to established g-estimation method for longitudinal de-mediation

**Figure S8.** Difference (in % of the SD of the estimates of the established g-estimation method for longitudinal measurements) between the SD of the estimates of the averaging g-estimation method and the SD of the estimates of the established g-estimation method for longitudinal measurements, in sets of 10.000 simulated trials per combination of parameters.

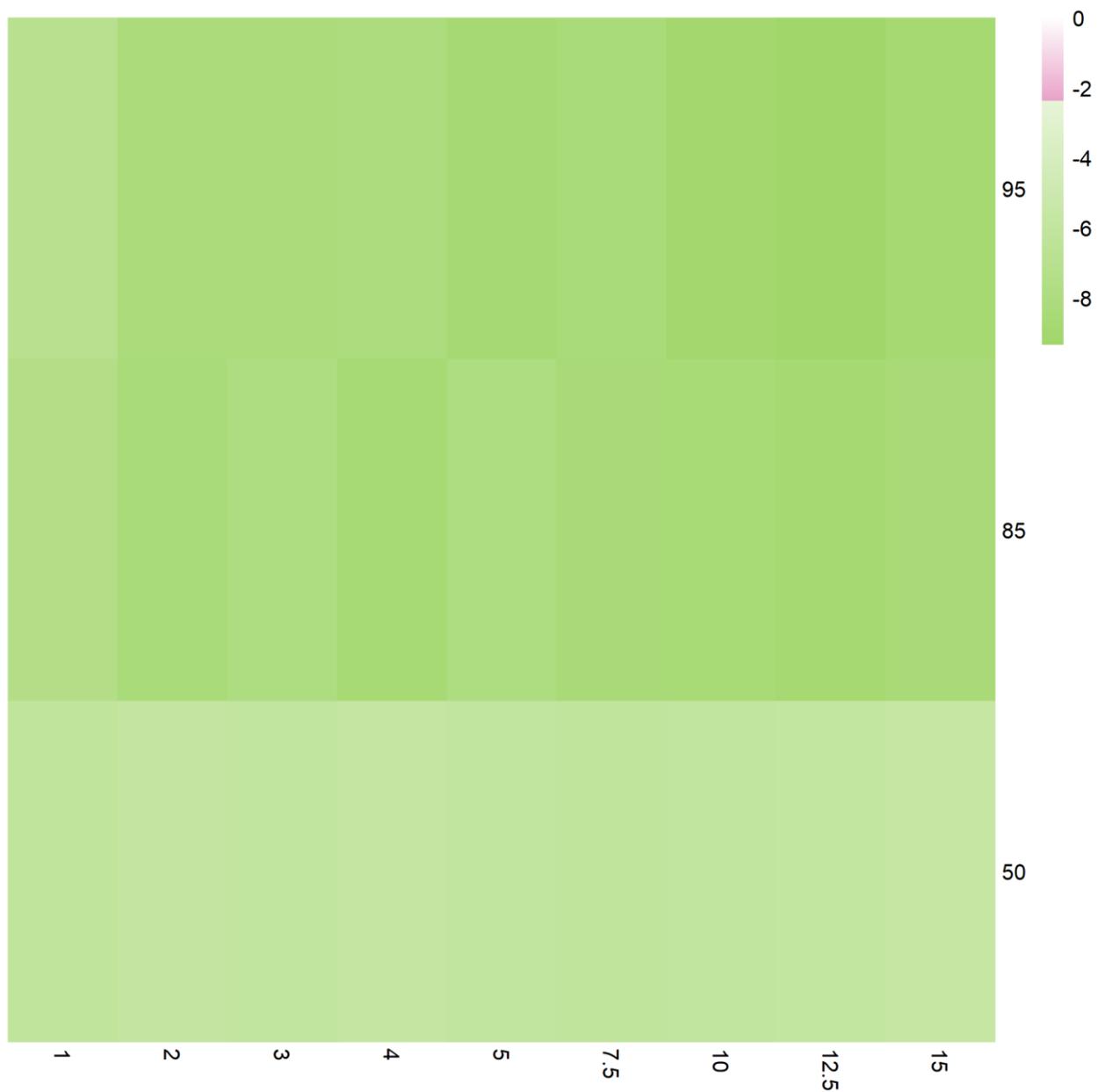

**Figure S9.** Difference (in % of the SD of the estimates of the established g-estimation method for longitudinal measurements) between the SD of the estimates of the averaging $Y_2^{\square}$ g-estimation method and the SD of the estimates of the established g-estimation method for longitudinal measurements, in sets of 10.000 simulated trials per combination of parameters.

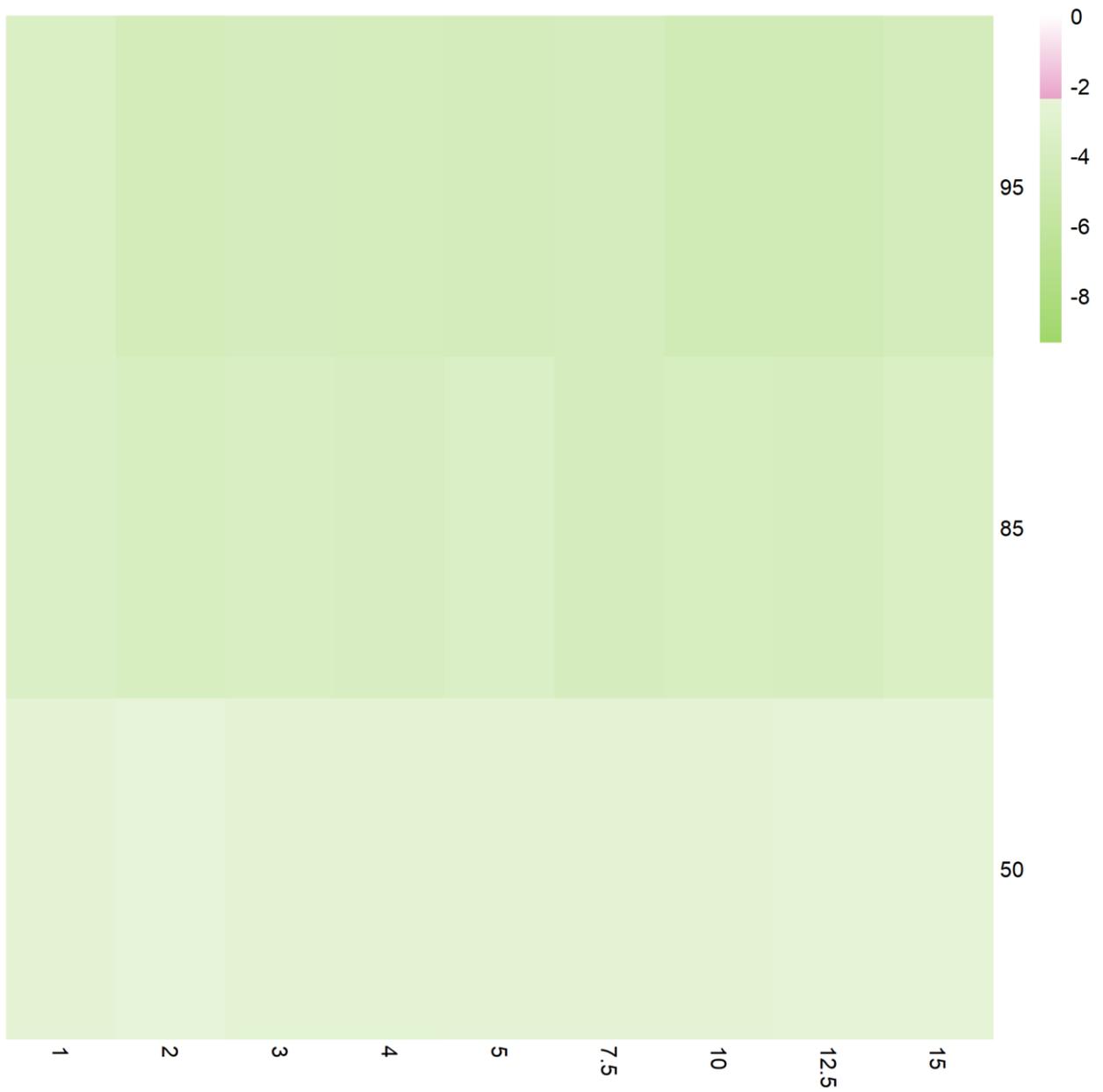

**Figure S10.** Difference (in % of the SD of the estimates of established g-estimation method for longitudinal measurements) between the SD of the estimates of the iterative g-estimation method and the SD of the estimates of the established g-estimation method for longitudinal measurements, in sets of 10.000 simulated trials per combination of parameters.

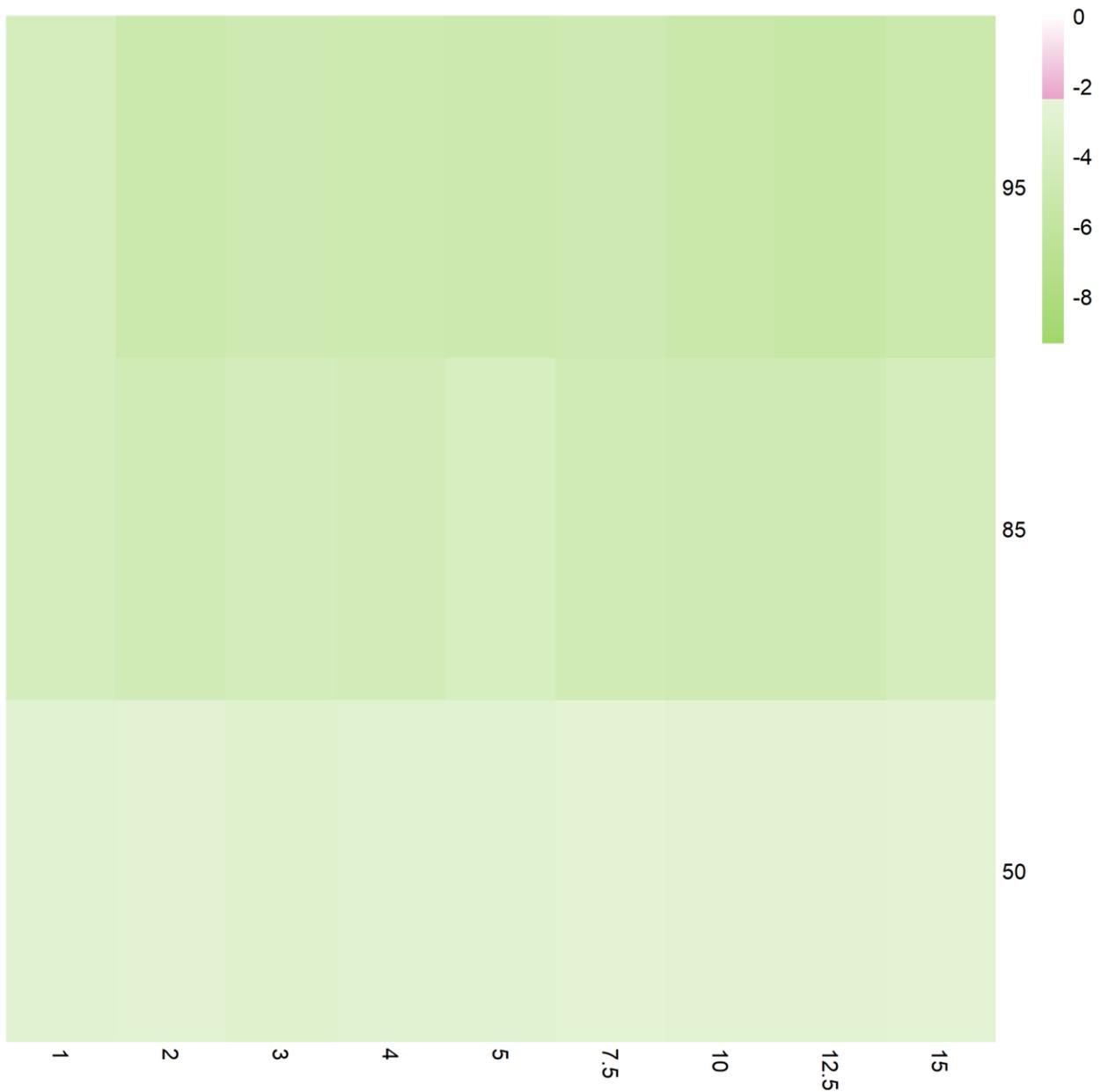

## 6.2.2. Standard deviation compared to a linear model based on the unaffected values Y_2*

**Figure S11.** Difference (in % of the SD of the estimates of linear model based on the unaffected values $Y_2^*$) between the SD of the estimates of the averaging g-estimation method and the SD of the estimates of the linear model based on the unaffected values $Y_2^*$, in sets of 10.000 simulated trials per combination of parameters.

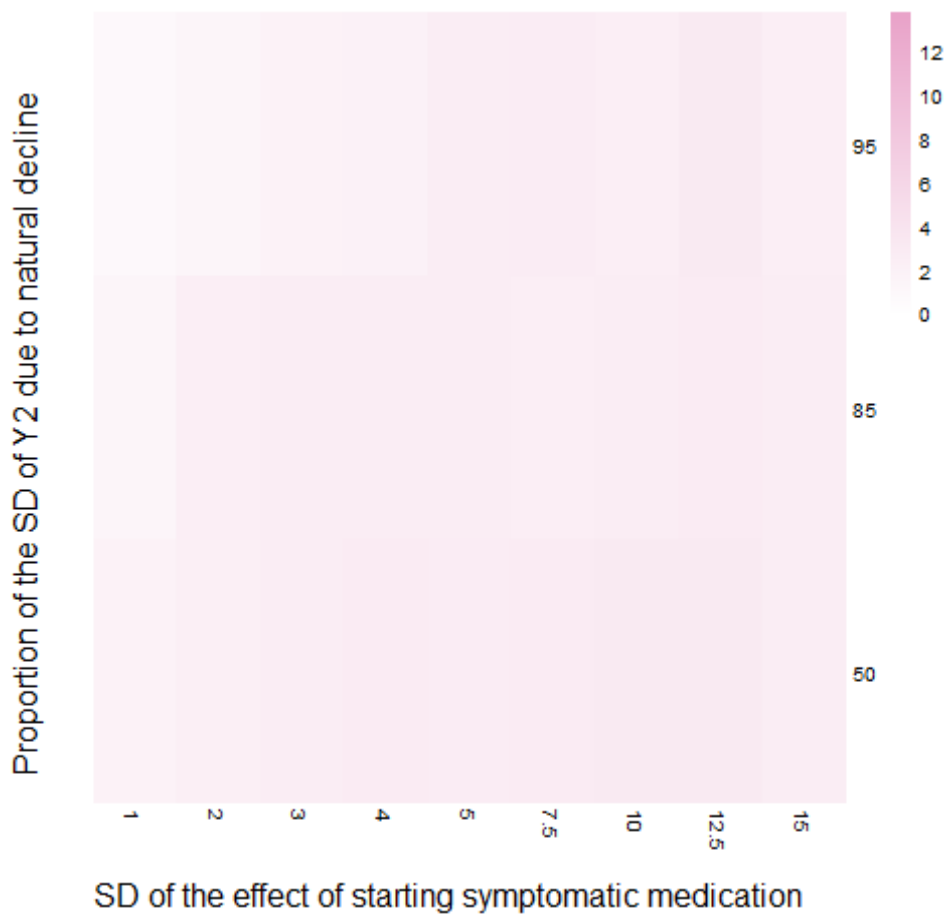

**Figure S12.** Difference (in % of the SD of the estimates of linear model based on the unaffected values $Y_2^*$) between the SD of the estimates of the averaging Y2 g-estimation method and the SD of the estimates of the linear model based on the unaffected values $Y_2^*$, in sets of 10.000 simulated trials per combination of parameters.

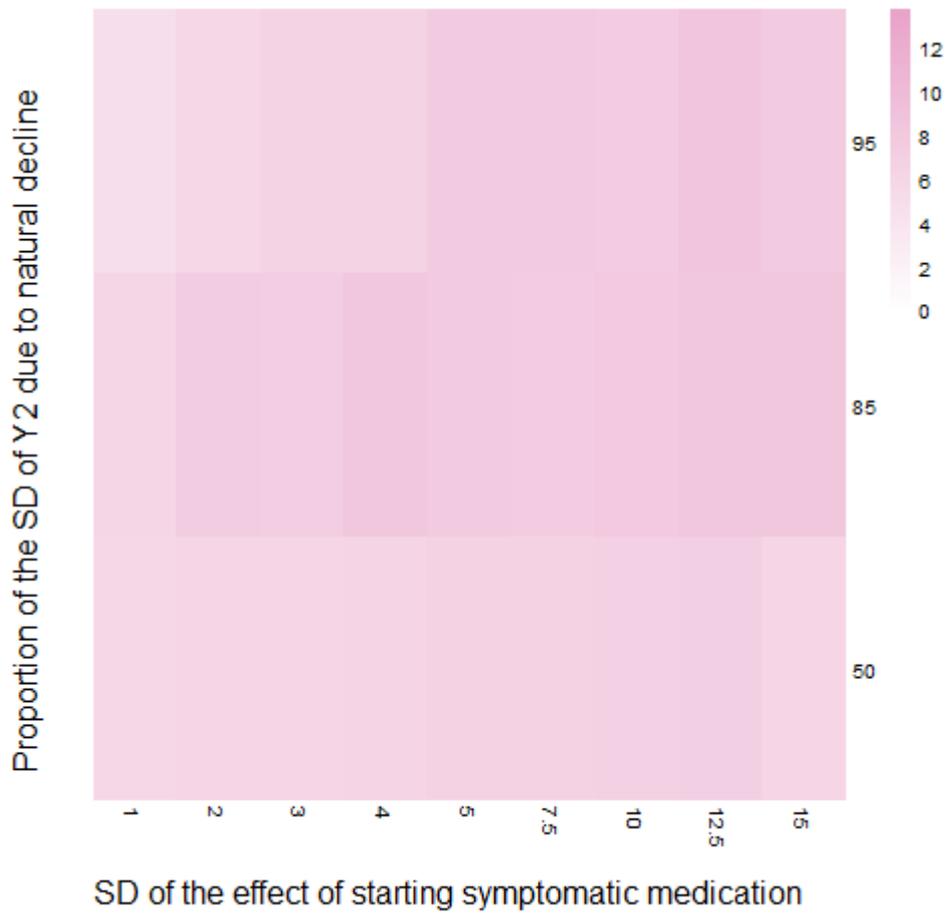

**Figure S13.** Difference (in % of the SD of the estimates of linear model based on the unaffected values $Y_2^*$) between the SD of the estimates of the iterative g-estimation method and the SD of the estimates of the linear model based on the unaffected values $Y_2^*$, in sets of 10.000 simulated trials per combination of parameters.

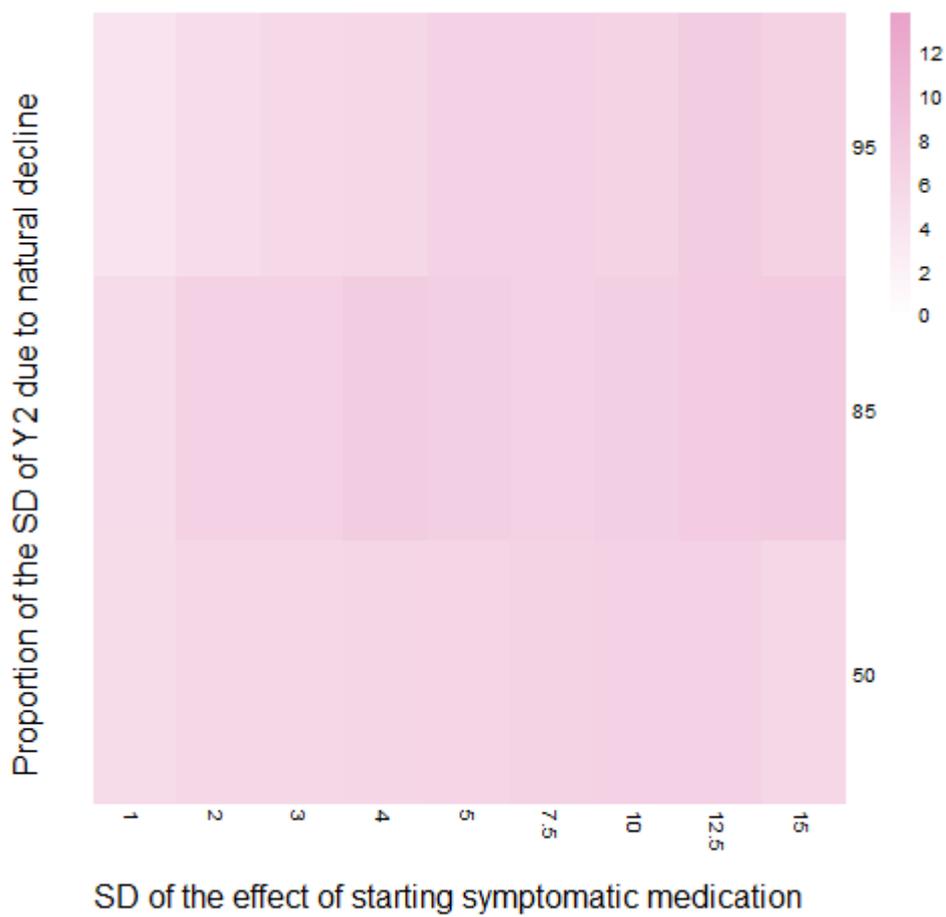

**Figure S14.** Difference (in % of the SD of the estimates of linear model based on the unaffected values $Y_2^*$) between the SD of the estimates of the established g-estimation method for longitudinal measurements and the SD of the estimates of the linear model based on the unaffected values $Y_2^*$, in sets of 10.000 simulated trials per combination of parameters.

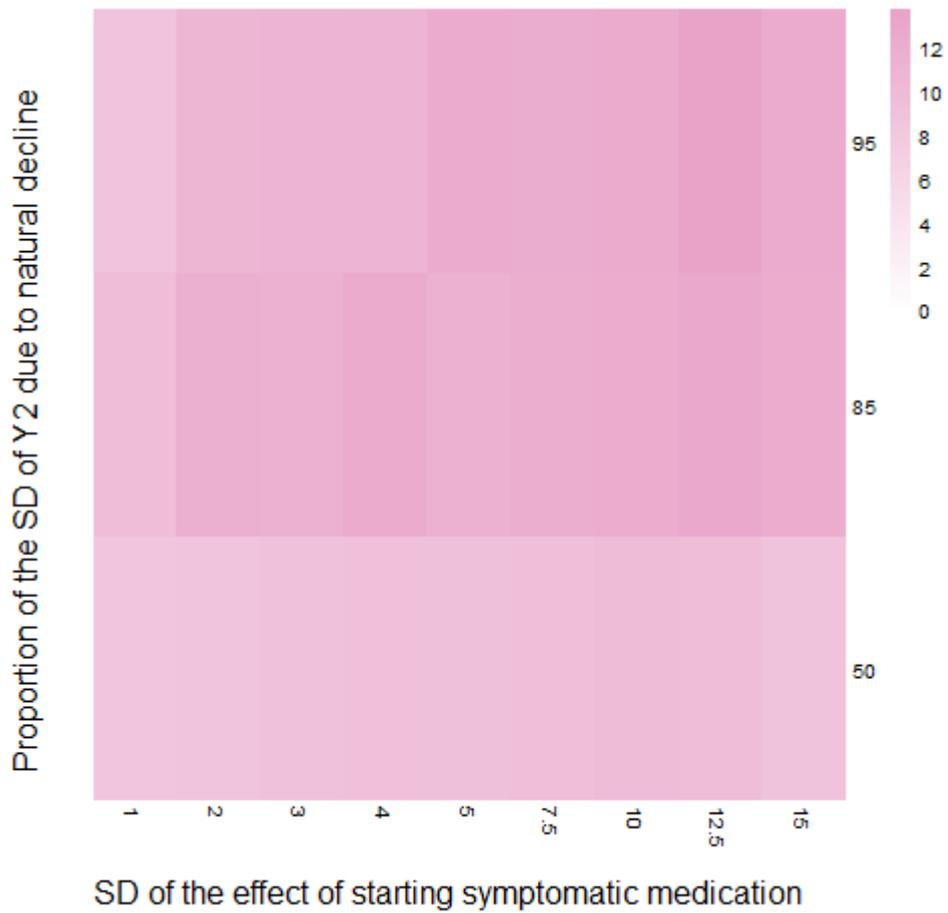

### 6.3. Bias sensitivity analysis

**Table S1.** Bias of the estimates of the established g-estimation method for longitudinal measurements in sets of 10.000 simulated trials per combination of parameters.

|        | SD of the effect of starting Symptomatic treatment | | | | | | | | |
|--------|----------|----------|----------|----------|----------|----------|------------|---------|---------|
| SD(Y2) | 1        | 2        | 3        | 4        | 5        | 7.5      | 10         | 12.5    | 15      |
| 10     | 0.041364 | 0.040568 | 0.046047 | 0.038528 | 0.043822 | 0.04911  | 0.04241083 | 0.04415 | 0.04344 |
| 13.5   | 0.07587  | 0.0922   | 0,069979 | 0.07102  | 0.09484  | 0.080314 | 0.086406   | 0.08808 | 0.08585 |

**Table S1.** Bias of Modification 1, Averaging the de-mediation on $Y_t$, in sets of 10.000 simulated trials per combination of parameters.

|        | SD of the effect of starting Symptomatic treatment | | | | | | | | |
|--------|----------|----------|----------|----------|----------|----------|----------|----------|----------|
| SD(Y2) | 1        | 2        | 3        | 4        | 5        | 7.5      | 10       | 12.5     | 15       |
| 10     | 0.008855 | 0.006154 | 0.011055 | 0.009956 | 0.009689 | 0.01448  | 0.012561 | 0.010299 | 0.013552 |
| 13.5   | 0.017696 | 0.024522 | 0.023337 | 0.01595  | 0.022381 | 0.023978 | 0.024447 | 0.02699  | 0.026458 |

**Table S1.** Bias of Modification 2, Averaging the de-mediation effects on $Y_2$, in sets of 10.000 simulated trials per combination of parameters.

|        | SD of the effect of starting Symptomatic treatment | | | | | | | | |
|--------|----------|---------|---------|---------|----------|---------|----------|----------|----------|
| SD(Y2) | 1        | 2       | 3       | 4       | 5        | 7.5     | 10       | 12.5     | 15       |
| 10     | 0.00893  | 0.00658 | 0.01188 | 0.00893 | 0.01418  | 0.01785 | 0.010732 | 0.015278 | 0.013071 |
| 13.5   | 0.025441 | 0.03573 | 0.03590 | 0.01833 | 0.036432 | 0.02698 | 0.034466 | 0.036518 | 0.03699  |

**Table S1.** Bias of Modification 3, Iterative averaging the de-mediation effects on $Y_2$, in sets of 10.000 simulated trials per combination of parameters.

|        | SD of the effect of starting Symptomatic treatment | | | | | | | | |
|--------|---------|----------|----------|----------|---------|----------|----------|----------|----------|
| SD(Y2) | 1       | 2        | 3        | 4        | 5       | 7.5      | 10       | 12.5     | 15       |
| 10     | 0.00617 | 0.001657 | 0.007207 | 0.006095 | 0.01159 | 0.01429  | 0.00999  | 0.01399  | 0.009635 |
| 13.5   | 0.02339 | 0.033142 | 0.035259 | 0.01558  | 0.03161 | 0.024232 | 0.031114 | 0.035693 | 0.035865 |

## 6.4. Investigation of a deterministic initiation mechanism for symptomatic treatment

**Table S5** Results of the de-mediation approaches under a DGM with a deterministic occurrence of the IE under the alternative hypothesis

|  | Empirical Bias* | Empirical standard deviation of the effect estimate |
|---|---|---|
| Established g-estimation approach for longitudinal de-mediation | 0.138 | 1.818 |
| Modification 1 Averaging the de-mediation on $Y_t$ | 0.085 | 1.790 |
| Modification 2 Averaging the de-mediation effects on $Y_2$ | 0.091 | 1.804 |
| Modification 3 Iterative averaging the de-mediation effects on $Y_2$ | 0.093 | 1.800 |

* The expected value of the effect (active treatment – placebo) under the alternative hypothesis is 0 on the ADAS-Cog 13 scale, where lower values correspond to a less sever disease. The bias is calculated as model estimate – expected value. A positive bias corresponds to an underestimation of the effect.

### 6.5. DAG code for DAGitty

```
dag {
bb="0,0,1,1"
decline [latent,pos="0.304,0.196"]
prognosis [latent,pos="0.116,0.194"]
sym05 [adjusted,pos="0.276,0.765"]
sym1 [adjusted,pos="0.411,0.768"]
sym15 [exposure,pos="0.541,0.758"]
treat [adjusted,pos="0.117,0.363"]
y0 [pos="0.212,0.436"]
y05 [pos="0.328,0.442"]
y1 [pos="0.421,0.572"]
y1.5 [adjusted,pos="0.542,0.585"]
y2 [outcome,pos="0.702,0.600"]
ystar1 [latent,pos="0.423,0.422"]
ystar15 [latent,pos="0.549,0.426"]
ystar2 [latent,pos="0.701,0.431"]
decline -> y05
decline -> ystar1
decline -> ystar15
decline -> ystar2
prognosis -> decline
prognosis -> y0
sym05 -> sym1
sym05 -> sym15 [pos="0.373,0.905"]
sym05 -> y1
sym05 -> y1.5
sym05 -> y2 [pos="0.648,0.925"]
sym1 -> sym15
sym1 -> y1.5
sym1 -> y2
sym15 -> y2
```

```
treat -> decline
y0 -> y05
y05 -> sym05
y05 -> ystar1
y1 -> sym1
y1.5 -> sym15
ystar1 -> y1
ystar1 -> ystar15
ystar15 -> y1.5
ystar15 -> ystar2
ystar2 -> y2
}
```